\documentclass[a4paper,11pt]{article}
\pdfoutput=1 

\usepackage{jheppub} 

\usepackage[T1]{fontenc} 
\usepackage{multirow}

\title{\boldmath Kink excitation spectra in the (1+1)-dimensional $\varphi^8$ model}

\author[a,b,1]{Vakhid A. Gani,\note{Corresponding author.}}
\author[b,c]{Vadim Lensky,}
\author[b,d]{Mariya A. Lizunova}

\affiliation[a]{Department of Mathematics, National Research Nuclear University MEPhI\\ (Moscow Engineering Physics Institute), 115409 Moscow, Russia}
\affiliation[b]{Theory Department, National Research Center Kurchatov Institute, Institute for Theoretical and Experimental Physics, 117218 Moscow, Russia}
\affiliation[c]{Department of Elementary Particle Physics, National Research Nuclear University MEPhI\\ (Moscow Engineering Physics Institute), 115409 Moscow, Russia}
\affiliation[d]{Department of Theoretical Nuclear Physics, National Research Nuclear University MEPhI\\ (Moscow Engineering Physics Institute), 115409 Moscow, Russia}

\emailAdd{vagani@mephi.ru}
\emailAdd{lensky@itep.ru}
\emailAdd{mary.lizunova@gmail.com}

\abstract{
We study excitation spectra of BPS-saturated topological
solutions --- the kinks --- of the $\varphi^8$ scalar field model
in $(1+1)$ dimensions, for three different choices of the model
parameters. We demonstrate that some of these kinks have
a vibrational mode, apart from the trivial zero (translational) excitation.
One of the considered kinks is shown to have three vibrational modes.
We perform a numerical calculation of the kink-kink scattering
in one of the considered variants of the $\varphi^8$ model,
and find the critical collision velocity $v_{\scriptsize \mbox{cr}}$
that separates the different collision regimes:
inelastic bounce of the kinks at $v_{\scriptsize \mbox{in}}\ge v_{\scriptsize \mbox{cr}}$, and capture at $v_{\scriptsize \mbox{in}}<v_{\scriptsize \mbox{cr}}$. We also observe escape windows
at some values of $v_{\scriptsize \mbox{in}}<v_{\scriptsize \mbox{cr}}$ where the kinks
escape to infinity after bouncing off each other two or more times. We analyse the features
of these windows and discuss their relation to the resonant energy exchange between the
translational and the vibrational excitations of the colliding kinks.
}
\keywords {kink, soliton, topological defect, domain wall, kink scattering, effective field theories, field theories in lower dimensions}

\makeatletter
\def\@fpheader{\relax}
\makeatother
\begin{document} 

\maketitle

\flushbottom

\section{\label{sec:level1} Introduction and motivation}

Topological defects in $(1+1)$ space-time dimensions have been a subject of active research~\cite{vilenkin01,manton,aek01}.
Properties of these defects, their interaction with other defects and impurities
are widely used to model various phenomena in condensed matter physics \cite{Alfimov:2014wta}.
For instance, field models considering a real scalar field with polynomial self-interaction
are routinely employed to model phase transitions. The most widely known
example of such a model is probably the $\varphi^4$ field theory.
It has a topological solution --- the kink --- that can be used to describe a system
undergoing a second order phase transition. More complex processes, when a series of consecutive phase
transitions is to be modelled, make necessary the use of polynomial self-interactions with the degree higher than four.

Field models that can develop topological solutions are also very important in the context of classical and quantum field theory, high-energy physics, cosmology, as well as hadron and nuclear physics \cite{Ahlqvist:2014uha,poltis1,poltis2,PhysRevD.38.3525,foster,foster02}. In this relation,
one has also to mention the progress in the research of properties of strings, vortices and
monopoles~\cite{vilenkin01,kibble01,kibble02,weigel02,weigel03,weigel04,palvelev,dziarmaga,myers,sutcliffe,ward,pitaevskii,loginov01,loginov02}.

To stress the importance of studying topological defects in $(1+1)$ dimensions, we note that,
while being sufficiently easily treatable, they can correctly describe structures in higher
dimensions, such as, for instance, a smooth domain wall in $(3+1)$ dimensions whose profile can be
described by a $(1+1)$-dimensional kink.

The existence of eigenmodes in the excitation spectrum of a solitary wave
is related to its stability against small perturbations; these modes can also affect the interaction
of the solitary waves with one another and with external objects. A soliton in a translationally
invariant model can always be shifted by a constant vector, which means that there always exists
the zero (translational) mode. Excitations with higher energies --- vibrational modes --- can
give rise to a rich collection of resonance phenomena in kink-(anti)kink collisions,
as well as in the scattering of kinks off impurities~\cite{krusch01,popov01,popov02,kivshar,BeKu,gumerov,saad01,saad02,saad03,oliveira01,oliveira02}.

This article deals with topological solitons of the $(1+1)$-dimensional $\varphi^8$ field theory.
This model has been employed, in particular, to model massless mesons with
self-interaction~\cite{lohe}, and to describe isostructural phase transitions~\cite{pavlov}.
We study for the first time the excitation spectra of the kinks that occur in this model, for three
different choices of the model's self-interaction. We show that some of
the studied kinks have a vibrational excitation mode. To illustrate how
the excitation spectrum of a single kink affects the latter's interaction with other kinks
(and other spatial defects), we select one of the kinks that have a vibrational excitation
and study its collisions with the corresponding antikink.

Our study is structured as follows. Section~\ref{sec:level2} provides a brief introduction into 
general properties of static solutions with finite energy in $(1+1)$ dimensions.
In Section~\ref{sec:level3}, we deal with the kinks of the $\varphi^8$ model (with three different
choices of the model parameters) and perform a study of their excitation spectra.
Section~\ref{sec:level4} presents the results of our numerical study of collisions between
one of the kinks that are considered in Section~\ref{sec:level3} and the corresponding antikink,
and a discussion of the observed resonance phenomena.
Section~\ref{sec:level5} delivers an outlook and the conclusion.

\section{\label{sec:level2} Kinks in $(1+1)$ dimensions}

We consider a field-theory system with a single real scalar field $\varphi(t,x)$ in one spatial
and one temporal dimensions, with the Lagrangian
\begin{equation}
	\mathcal{L}=\frac{1}{2} \left( \partial_{\mu}\varphi \right) ^2-V(\varphi),~~\mu=0,1,	
	\label{eq:largang}
\end{equation}
where the self-interaction of the field $\varphi$, $V(\varphi)$, is assumed to be bounded from
below and hence can be thought of as a non-negative function of $\varphi$.
The energy functional, corresponding to Eq.~(\ref{eq:largang}), is
\begin{equation}
	E[\varphi]=\int\limits_{-\infty}^{\infty}\left[\frac{1}{2} \left( \frac{\partial\varphi}{\partial t} \right)^2+\frac{1}{2} \left( 
\frac{\partial\varphi}{\partial x} \right) ^2+V(\varphi)\right]dx.
    \label{eq:energ}
\end{equation}
The Lagrangian~(\ref{eq:largang}) yields the following equation of motion for the field $\varphi(t,x)$:
\begin{equation}
	\Box\varphi+\frac{dV}{d\varphi}=0,
	\label{eq:eom}
\end{equation}
where $\Box=\partial^2_t-\partial^2_x$ is the d'Alembertian. For a static configuration, $\varphi=\varphi(x)$, this becomes
\begin{equation}
	\frac{d^2\varphi}{dx^2}=\frac{dV}{d\varphi},
	\label{eq:steom}
\end{equation}
which can be transformed into
\begin{equation}
	\frac{d\varphi}{dx}=\pm\sqrt{2V(\varphi)}
	\label{eq:steomv}
\end{equation}
or further into
\begin{equation}
	\frac{d\varphi}{dx}=\pm\frac{dW}{d\varphi},
	\label{eq:steomw}
\end{equation}
where the superpotential $W(\varphi)$ is related with $V(\varphi)$ via
\begin{equation}
	V(\varphi)=\frac{1}{2}\left(\frac{dW}{d\varphi}\right)^2.
	\label{eq:dwdfi}
\end{equation}

If the potential $V(\varphi)$ has two or more minima $\overline{\varphi}_1$, $\overline{\varphi}_2,\ \dots$, yielding the same minimal value $V(\overline{\varphi}_i)=0,$ $i=1,2,\dots$,
the energy of a static solution $\varphi=\varphi(x)$ can be written as
\begin{equation}
	E=E_{\scriptsize \mbox{BPS}}+\frac{1}{2}\int\limits_{-\infty}^{\infty}\left(\frac{d\varphi}{dx}\mp\frac{dW}{d\varphi}\right)^2dx,
	\label{eq:enfull}
\end{equation}
where
\begin{equation}
    E_{\scriptsize \mbox{BPS}}=|W[\varphi(+\infty)]-W[\varphi(-\infty)]|
    \label{eq:enbps}
\end{equation}
is the energy of the BPS-saturated solution~\cite{manton,bps1,bps2}, where
the static field $\varphi(x)$ fulfils Eq.~(\ref{eq:steomw}) and therefore
the integrand in Eq.~(\ref{eq:enfull}) turns into zero. The BPS-saturated solution
therefore has the smallest energy among all the static solutions interpolating between
two given adjacent minima of the potential,
\begin{equation*}
\varphi(-\infty)=\lim_{x \to -\infty}\varphi(x),~
\varphi(+\infty)=\lim_{x \to +\infty}\varphi(x).
\end{equation*}
Note that the static solution $\varphi(x)$ has to converge sufficiently
fast to one of the minima of the potential in order that the energy be finite,
\begin{equation}
    \lim_{x \to -\infty}\varphi(x)=\overline{\varphi}_i,~
    \lim_{x \to +\infty}\varphi(x)=\overline{\varphi}_j
    \label{eq:asympt}
\end{equation}
(this is also true if the field depends on time).
As usual, solutions where $\overline{\varphi}_i=\overline{\varphi}_j$ are called non-topological,
whereas $\overline{\varphi}_i\neq\overline{\varphi}_j$ corresponds to a topological solution.
We will call the family of all solutions with identical
spatial asymptotics a ``topological sector'', with the corresponding notation, e.g.,
$(\overline{\varphi}_i,\overline{\varphi}_j)$ for the 
topological sector with the asymptotics of Eq.~(\ref{eq:asympt}).

The field theory being Lorentz-invariant, a static solution of Eq.~(\ref{eq:steom}) or
Eq.~(\ref{eq:steomw}) can be boosted to produce a soliton moving with a constant velocity $v$,
with the energy given by
\[
E=\frac{M}{\sqrt{1-v^2}},
\]
with $M$ being the energy of the static kink. We reserve the term ``kink'' (``antikink'') for
topological BPS-saturated solutions that connect any two
adjacent minima of the self-interaction potential, and their boosts.

In order to analyse the excitation spectrum of a static kink $\varphi_{\scriptsize \mbox{k}}(x)$,
we add to it a small perturbation,
\begin{equation}
    \varphi(t,x)=\varphi_{\scriptsize \mbox{k}}(x)+\delta\varphi(t,x), \quad ||\delta\varphi||\ll||\varphi_{\scriptsize \mbox{k}}||,
    \label{eq:phiexpan}
\end{equation}
and, taking in the equation of motion~(\ref{eq:eom}) terms linear in $\delta\varphi(t,x)$, obtain:
\begin{equation}
    \frac{\partial^2\delta\varphi}{\partial t^2}-\frac{\partial^2\delta\varphi}{\partial 
x^2}+\left.\frac{d^2V}{d\varphi^2}\right|_{\varphi_{\scriptsize \mbox{k}}(x)}\cdot\delta\varphi=0.
    \label{eq:eomexpan}
\end{equation}
Looking for $\delta\varphi(t,x)$ in the form
\begin{equation}
    \delta\varphi(t,x)=\psi(x)\cos(\omega t),
    \nonumber
    \label{eq:subexpan}
\end{equation}
we obtain from~(\ref{eq:eomexpan}) a boundary value problem
\begin{equation}
    \left[ -\frac{d^2}{dx^2}+\left.\frac{d^2V}{d\varphi^2}\right|_{\varphi_{\scriptsize \mbox{k}}(x)} \right]\psi(x)=\omega^2\psi(x),
    \label{eq:shredexpan}
\end{equation}
where $\psi(x)$ has to satisfy the usual Schr\"odinger-like conditions, i.e., it has to be smoothly
differentiable and square integrable over the real axis.
The similarity to the Schr\"odinger equation can be further exploited by denoting
\begin{equation}
    U(x)=\left.\frac{d^2V}{d\varphi^2}\right|_{\varphi_{\scriptsize \mbox{k}}(x)},
    \label{eq:uxexpan}
    \nonumber
\end{equation}
making Eq.~(\ref{eq:shredexpan}) an eigenvalue problem for the Hamiltonian
\begin{equation}
    \hat{H}=-\frac{d^2}{dx^2}+U(x).
    \label{eq:gamexpan}   
\end{equation}
It can easily be shown that the Hamiltonian~(\ref{eq:gamexpan}) always has a zero
eigenvalue, corresponding to the translational excitation mode.
To demonstrate this, take the derivative of Eq.~(\ref{eq:steom}) with respect to $x$,
which gives
\begin{equation}
    -\frac{d^2\varphi^{\prime}}{dx^2}+\frac{d^2V}{d\varphi^2}\cdot\varphi^{\prime}=0.
    \label{eq:steqexpan}     
\end{equation}
If we substitute $\varphi=\varphi_{\scriptsize \mbox{k}}(x)$, this equation becomes an identity
(as the kink is a solution of the equation of motion),
at the same time coinciding with Eq.~(\ref{eq:shredexpan}) if one selects $\omega=0$. Hence,
\begin{equation}
    \psi_0(x)=\varphi_{{\scriptsize \mbox{k}}}^{\prime}(x)
\label{eq:phiprime}
\end{equation}
is the eigenfunction of the Hamiltonian~(\ref{eq:gamexpan}), corresponding to the eigenvalue
$\omega=0$. The square integrability of $\psi_0(x)$ follows from the fact that the energy of the kink
is finite, cf.\ Eq.~(\ref{eq:energ}), therefore the eigenvalue $\omega=0$ belongs to the discrete part of the excitation spectrum.

\section{\label{sec:level3} The kinks of the $\varphi^8$ model and their excitation spectra}

The $\varphi^8$ model is described by the Lagrangian~(\ref{eq:largang}), where the potential is
a polynomial having the degree eight. The shape of the potential can vary depending on the
values of the polynomial coefficients. In turn, different shapes of the potential
can generate different sequences of phase transitions in condensed matter systems,
as shown in ref.~\cite{khare}. Our choice of the self-interaction parameters follows this reference;
the specific potentials that we use have two, three or four degenerate minima, and all are non-negative
functions of the field $\varphi$.

\subsection{\label{sec:level3A} Four degenerate minima}

\begin{figure}
\begin{center}
	\includegraphics[scale=0.5]{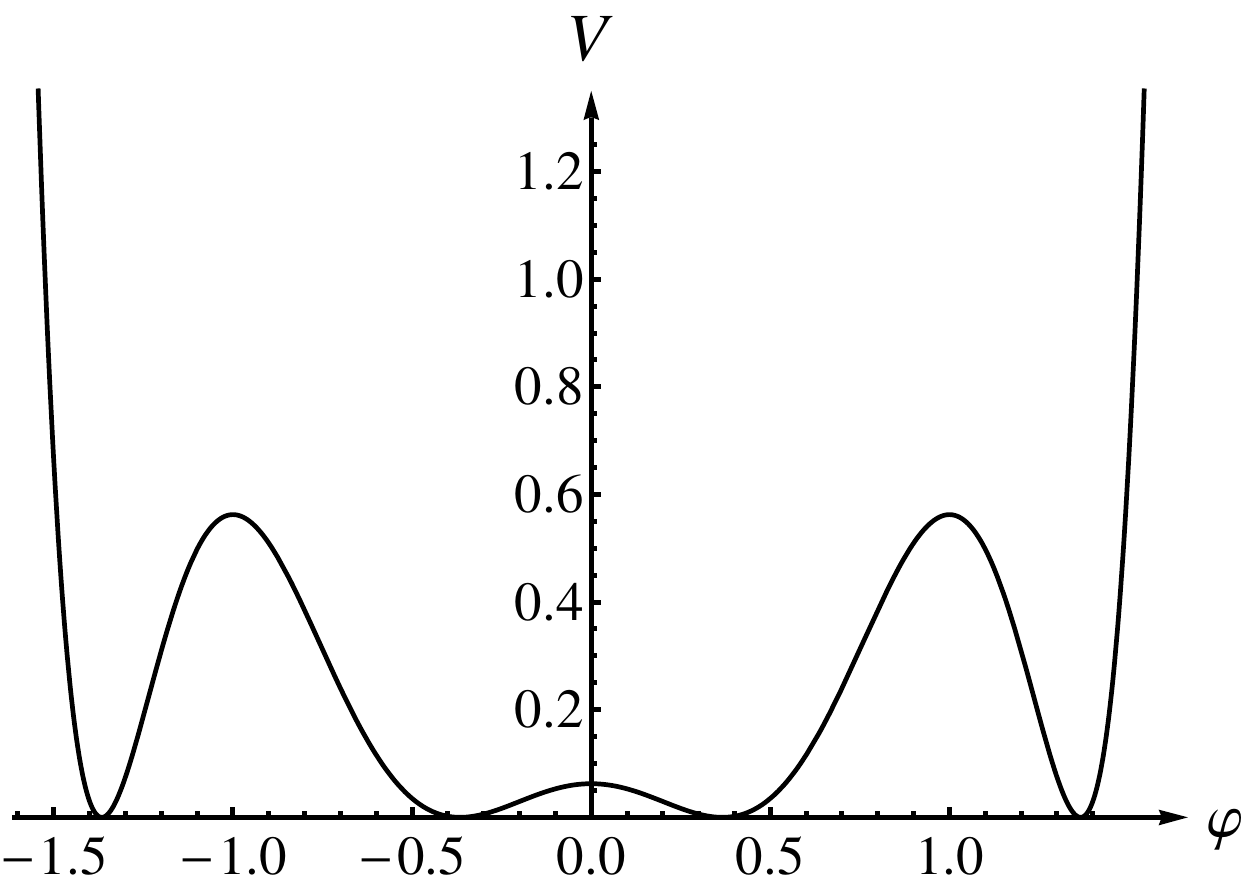}
\end{center}
	\caption{Plot of the potential with four degenerate minima~(\ref{eq:potphi8}).}
	\label{fig:potkink1}
\end{figure}

The shape of the potential can in this case be parameterised as
\begin{equation}
    V(\varphi)=\lambda^2(\varphi^2-a^2)^2(\varphi^2-b^2)^2,
    \label{eq:potphi8}
\end{equation}
where the constants are $0<a<b$, $\lambda>0$. The potential~(\ref{eq:potphi8})
has four degenerate minima $\overline{\varphi}_1=-b$, $\overline{\varphi}_2=-a$,
$\overline{\varphi}_3=a$, and $\overline{\varphi}_4=b$. Following ref.~\cite{khare},
we use 
\begin{equation}
    a=\displaystyle\frac{-1+\sqrt{3}}{2},~~~b=\displaystyle\frac{1+\sqrt{3}}{2}.
    \label{eq:constants1}
\end{equation}
We also set $\lambda=1$ in the numerical calculations, which amounts to measuring
the potential $V(\varphi)$ in units of $\lambda^2$, while $x$ and $t$ --- in units of $\lambda^{-1}$.
Fig.~\ref{fig:potkink1} shows the potential~(\ref{eq:potphi8}) corresponding to the above
choice of parameters.
For this potential, Eq.~(\ref{eq:steomv}) gives:
\begin{equation}
 \sqrt{2}\lambda x=\int\frac{d\varphi}{\sqrt{(\varphi^2-a^2)^2(\varphi^2-b^2)^2}}.
 \label{eq:eombogphi8a}
\end{equation}
Looking at Fig.~\ref{fig:potkink1}, one can deduce that the kinks that connect the neighbouring vacua
belong to either of the three topological sectors $(-b,-a)$, $(-a,a)$, or $(a,b)$.
For example, a static BPS-saturated solution of
Eq.~(\ref{eq:eombogphi8a}) interpolating between the vacua $\varphi=-a$ at $x\to -\infty$ and
$\varphi=a$ at $x\to +\infty$ belongs to the sector $(-a,a)$.
At the same time, there is also the corresponding antikink that connects the same vacua, but
$\varphi=a$ at $x\to -\infty$ and $\varphi=-a$ at $x\to +\infty$.
Formally, it belongs to the topological sector $(a,-a)$;
the distinction between kinks and antikinks is, however, just a matter of convention,
and we will not distinguish between the sectors $(-a,a)$ and $(a,-a)$ and will drop the prefix ``anti'',
unless the opposite is needed in order to avoid confusion or to explicitly identify the
field configuration in question.

The kinks corresponding to the three topological sectors above can be obtained from
Eq.~(\ref{eq:eombogphi8a}) as implicit functions of $x$~\cite{lohe,khare}. Below, we
examine their excitation spectra.

\subsubsection{\label{sec:level3A1} Topological sector $(-a,a)$}

\begin{figure}[h]
\begin{minipage}[h]{0.49\linewidth}
\center{\includegraphics[width=0.9\linewidth]{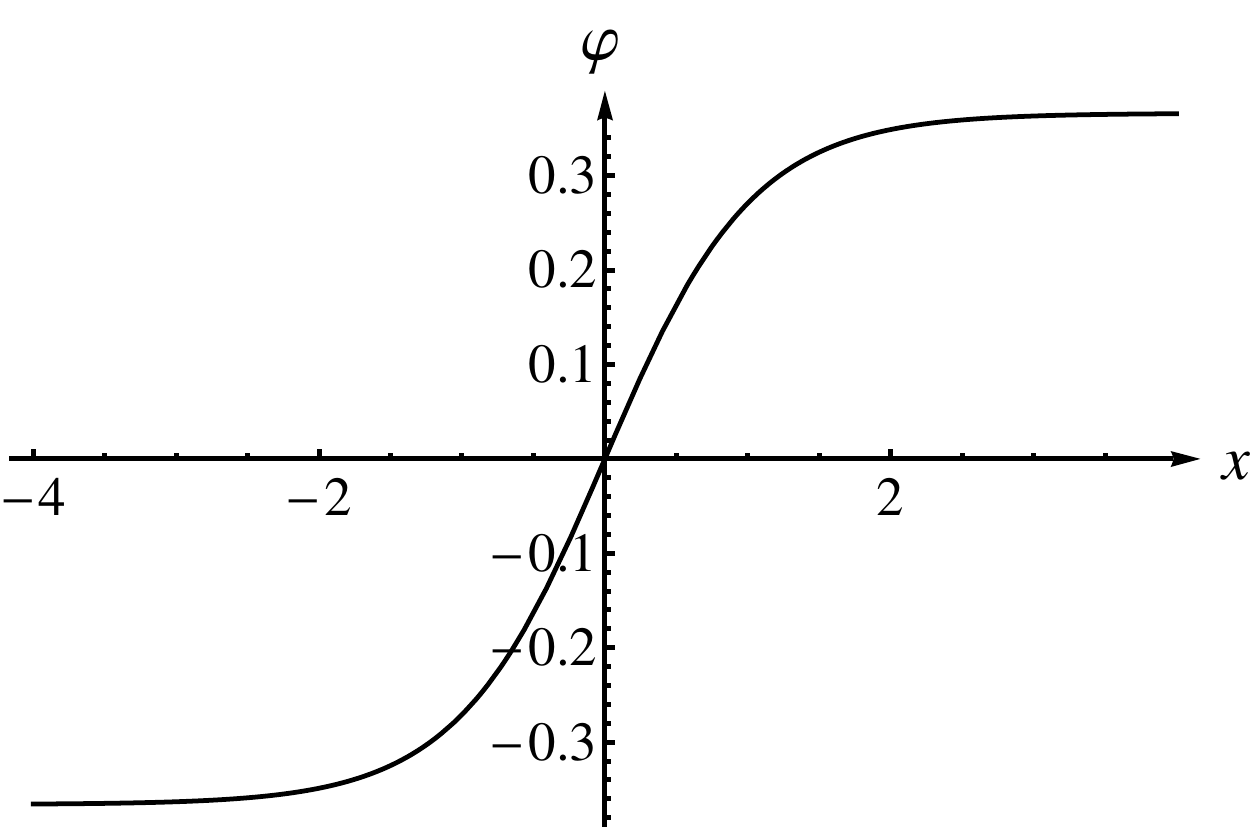}}
\end{minipage}
\hfill
\begin{minipage}[h]{0.49\linewidth}
\center{\includegraphics[width=0.9\linewidth]{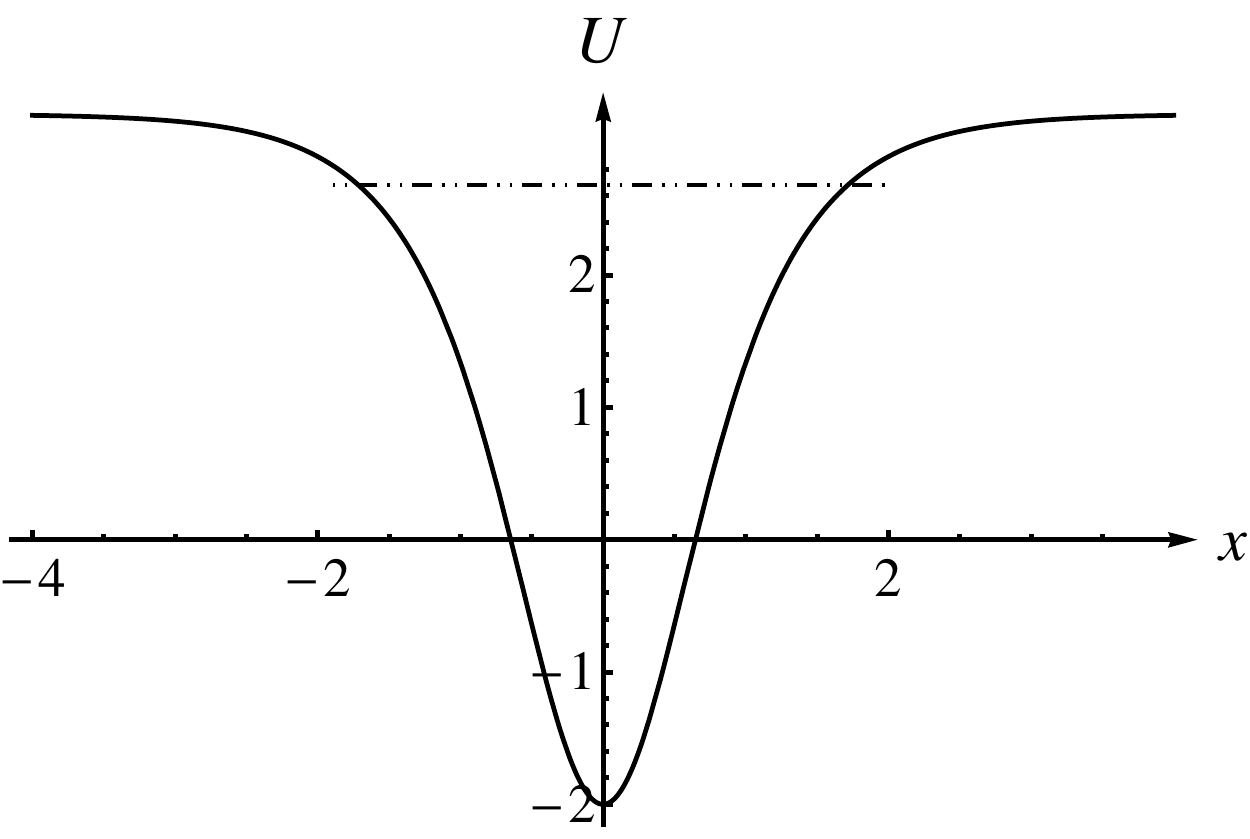}}
\end{minipage}
\caption{{\bf Left panel} --- the solution of Eq.~(\ref{eq:kinkph8a}) that connects the vacua
$-a$ and $a$. {\bf Right panel} --- the potential $U(x)$ corresponding to the kink shown in
the left panel. The dashed line shows the value of $\omega_1^2$.}
\label{fig:kink1}
\end{figure}

This kink is constrained by $|\varphi|<a$, which allows Eq.~(\ref{eq:eombogphi8a}) to be rewritten as
\[
 \sqrt{2}\lambda x=\int\frac{d\varphi}{(a^2-\varphi^2)(b^2-\varphi^2)}.
\]
The corresponding implicit solution is~\cite{lohe,khare}:
\begin{equation}
    e^{\mu x}=\frac{a+\varphi}{a-\varphi}\left(\frac{b-\varphi}{b+\varphi}\right)^{a/b},
    \label{eq:kinkph8a}
\end{equation}
where $\mu=2\sqrt{2}\lambda a(b^2-a^2)$. This equation can be solved for $\varphi(x)$ numerically,
and the plot of this kink is shown in Fig.~\ref{fig:kink1} (left panel).
The corresponding antikink can be obtained from Eq.~(\ref{eq:kinkph8a})
by substituting $x\to -x$:
\begin{equation}
    e^{-\mu x}=\frac{a+\varphi}{a-\varphi}\left(\frac{b-\varphi}{b+\varphi}\right)^{a/b}.
    \label{eq:kinkph8ak}
\end{equation}
Using Eq.~(\ref{eq:energ}) or Eq.~(\ref{eq:enbps}) gives the energy (mass) of the static
kink:
\[
    M_{(-a,a)}=\frac{4\sqrt{2}}{15}\lambda a^3(5b^2-a^2).
\]
We performed a numerical search of excitations of the kink~(\ref{eq:kinkph8a})
lying in the discrete part of the excitation spectrum. Fig.~\ref{fig:kink1} (right panel)
shows the corresponding potential $U(x)$ that enters the Schr\"odinger eigenvalue problem.
This problem was solved using the standard methods, namely, integrating Eq.~(\ref{eq:shredexpan}),
with the known asymptotic behaviour of its solutions at $x\to\pm \infty$, starting at a large
negative $x=x_l$ (the ``left'' solution) and a large positive $x=x_r$ (the ``right'' solution).
The two solutions were then matched at some point $\tilde{x}$ close to the spatial origin.
The specific choice of $\tilde{x}$ is not very important, for instance, one could take $\tilde{x}=0$,
however, it is convenient to take a small offset from the zero value --- this helps to avoid technical
issues when $U(x)$ is an even function of $x$
(in this case, excitation profiles $\psi(x)$ are either even or odd functions of $x$, and
the latter ones have a node at $x=0$).
We selected those values of $\omega$ at which the Wronskian of the ``left'' and the ``right''
solution, calculated at the matching point, turns to zero.

We found two eigenvalues, $\omega^2_0=-2\times 10^{-8}$, and $\omega^2_1 = 2.70491$.
The former value is just the translational mode, whose exact energy is $\omega_0=0$;
the deviation of our numerical result from zero thus provides an estimate
of accuracy. The latter value corresponds to the vibrational excitation, whose existence is
non-trivial and reflects itself in resonance phenomena occurring in kink-antikink collisions
in this topological sector, see Sec.~\ref{sec:level4}.

\subsubsection{\label{sec:level3A2} Topological sector $(-b,-a)$}

\begin{figure}[h]
\begin{minipage}[h]{0.49\linewidth}
\center{\includegraphics[width=0.9\linewidth]{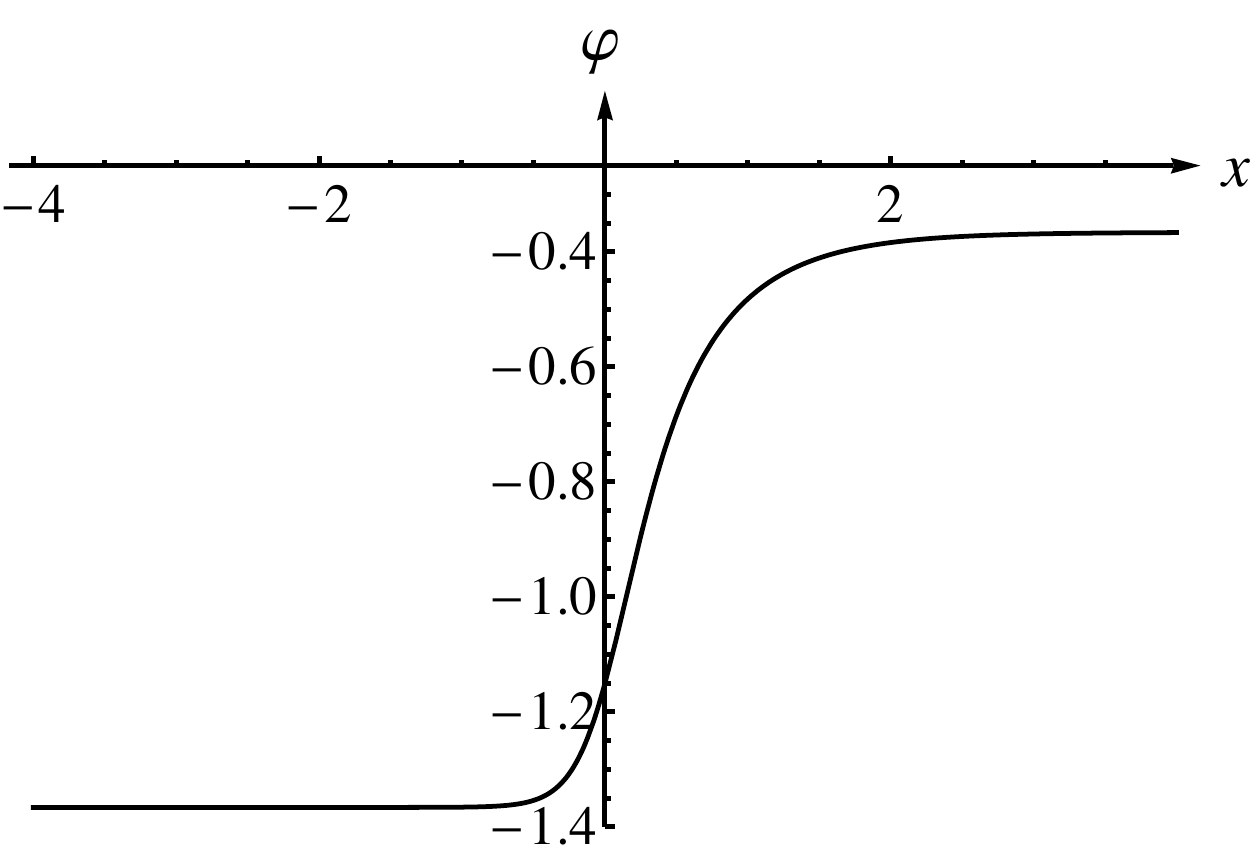} }
\end{minipage}
\hfill
\begin{minipage}[h]{0.49\linewidth}
\center{\includegraphics[width=0.9\linewidth]{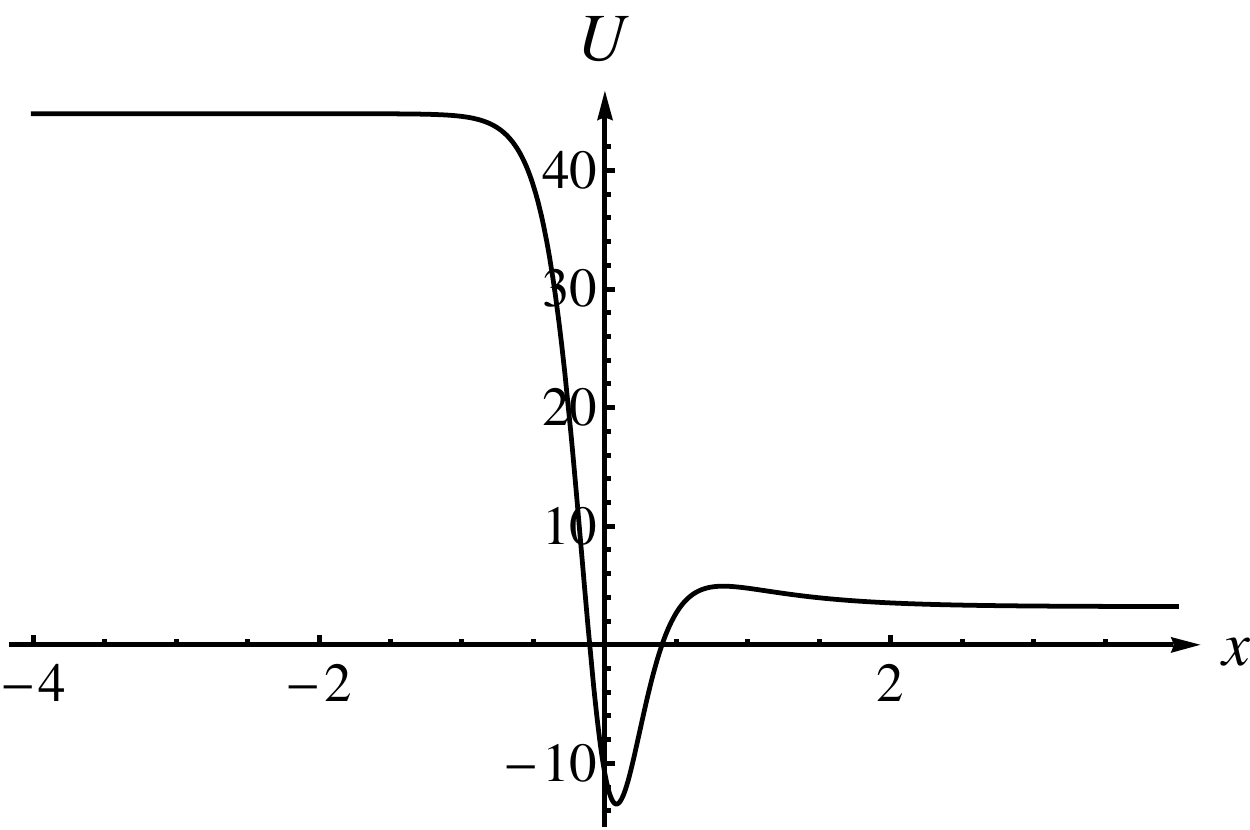}}
\end{minipage}
\caption{{\bf Left panel} --- the kink~(\ref{eq:kinkph8b}) that connects the vacua $-b$ and $-a$.
{\bf Right panel} --- the potential $U(x)$ corresponding to the kink shown in the left panel.}
\label{fig:kink2}
\end{figure}

In this sector $a<|\varphi|<b$, which turns (\ref{eq:eombogphi8a}) into
\[
 \sqrt{2}\lambda x=\int\frac{d\varphi}{(\varphi^2-a^2)(b^2-\varphi^2)}.
\]
Taking the integral results in
\begin{equation}
    e^{\mu x}=\frac{\varphi-a}{\varphi+a}\left(\frac{b+\varphi}{b-\varphi}\right)^{a/b},
    \label{eq:kinkph8b}
\end{equation}
where $\mu=2\sqrt{2}\lambda a(b^2-a^2)$; note that the kink that connects the vacua
$a$ and $b$ can be directly obtained from this expression.
The static kink (\ref{eq:kinkph8b}) has the mass
\[
    M_{(-b,-a)}=\frac{2\sqrt{2}}{15}\lambda (b-a)^3(a^2+3ab+b^2);
\]
its profile is shown in Fig.~\ref{fig:kink2} (left panel).

As opposed to the sector $(-a,a)$, we found the kink (\ref{eq:kinkph8b}) of the sector $(-b,-a)$
(which applies to that of the sector $(b,a)$, too) to only have the trivial translational
excitation, the numerical value being $\omega^2_0= -7\times 10^{-8}$.
The corresponding potential entering the Schr\"odinger equation is shown in the
right panel of Fig.~\ref{fig:kink2}.
It has to be noted that, even though the kinks considered in this subsection do not
have a vibrational excitation, a static kink and the corresponding static antikink located
close to each other can still have a vibrational mode.
Notice that even though such a configuration is not a solution of the equation of motion,
due to the nonlinear character of the field interaction, the introduced error typically falls
off exponentially with increasing separation between the two solitons,
which is why such Ans\"atze are routinely employed. The existence of such a ``collective'' vibrational mode can manifest itself as resonance phenomena in kink-antikink collisions.
Realisations of this mechanism have been studied, for instance, in refs.~\cite{dorey01,GaKuPRE}.

\subsection{\label{sec:level3B} Three degenerate minima}

\begin{figure}
\begin{center}
	\includegraphics[scale=0.5]{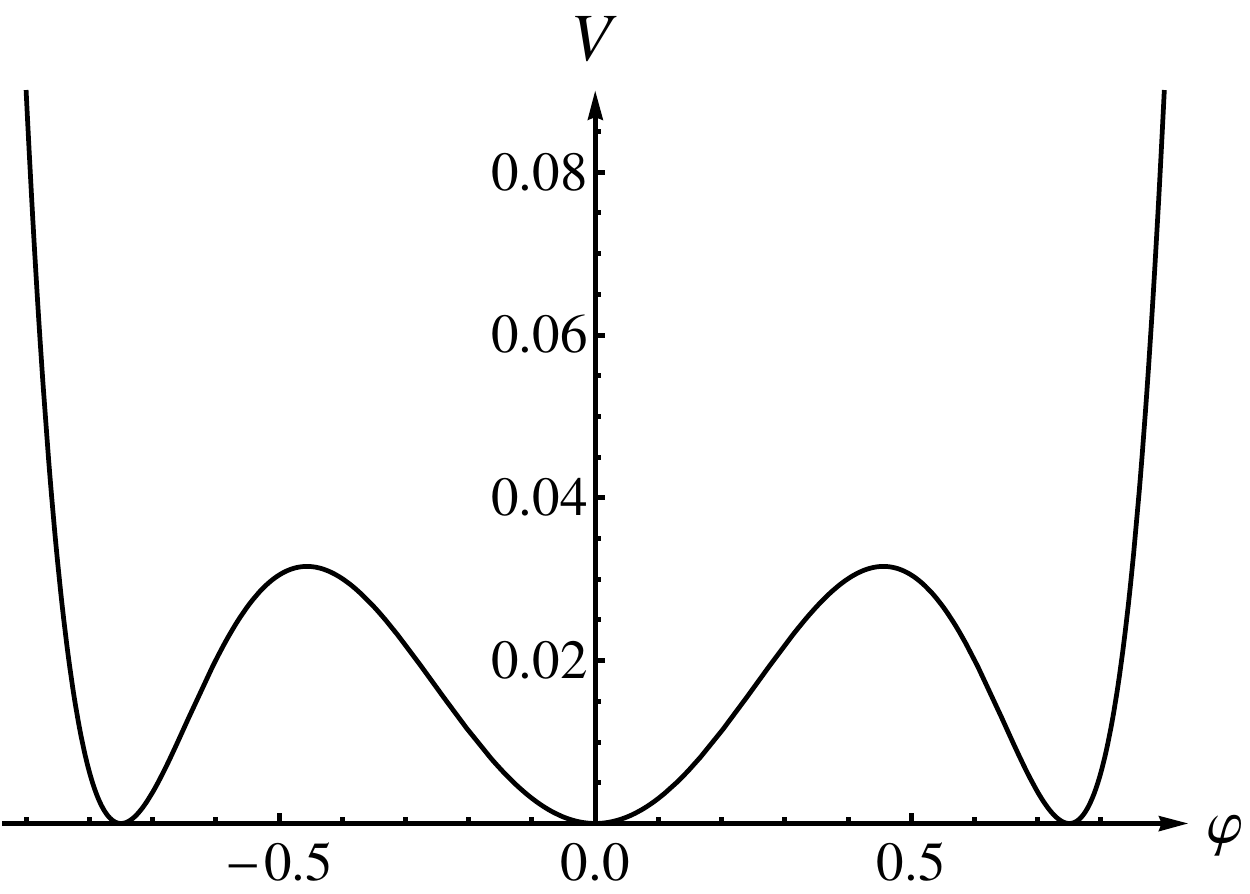}
\end{center}
	\caption{Plot of the potential with three degenerate minima~(\ref{eq:potphi8b}).}
	\label{fig:potkink2}
\end{figure}

\begin{figure}[h]
\begin{minipage}[h]{0.49\linewidth}
\center{\includegraphics[width=0.9\linewidth]{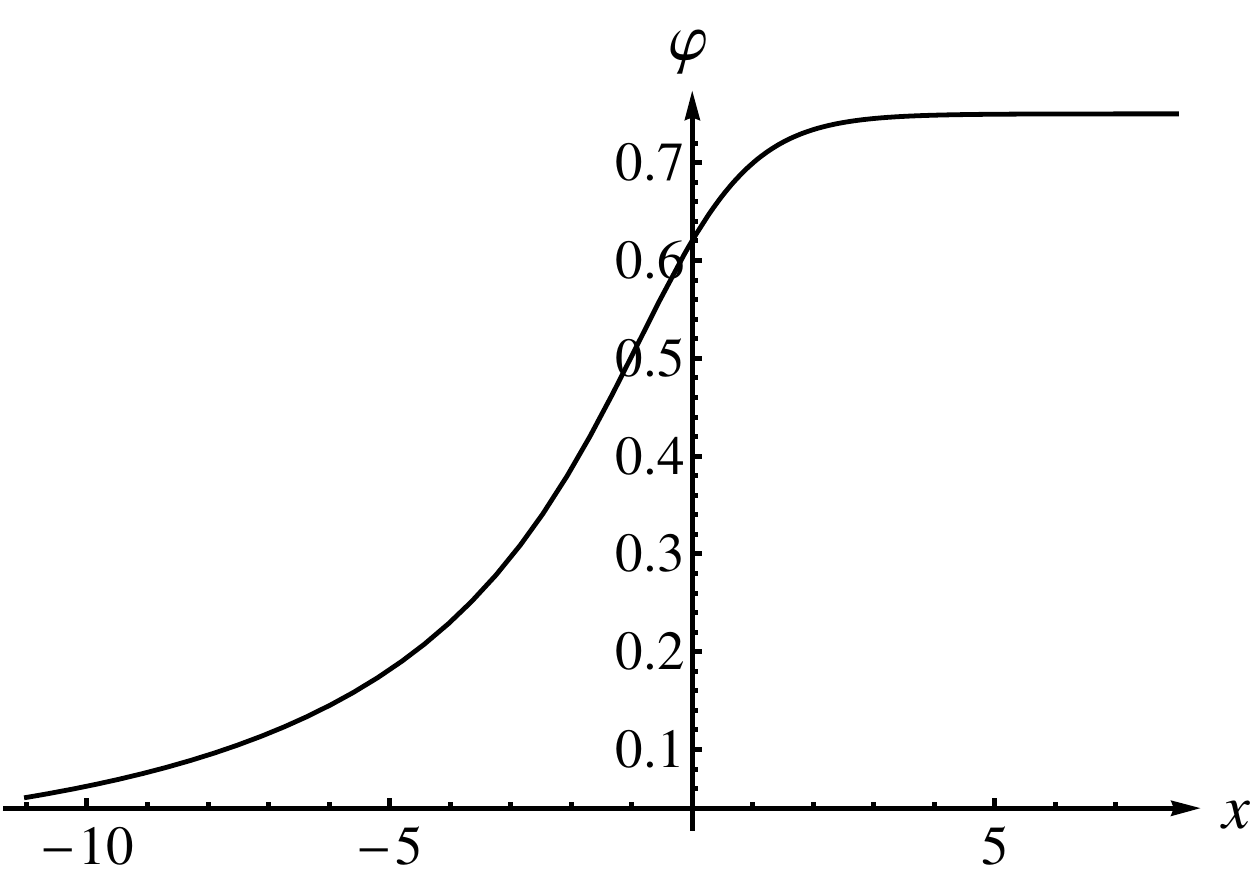}}
\end{minipage}
\hfill
\begin{minipage}[h]{0.49\linewidth}
\center{\includegraphics[width=0.9\linewidth]{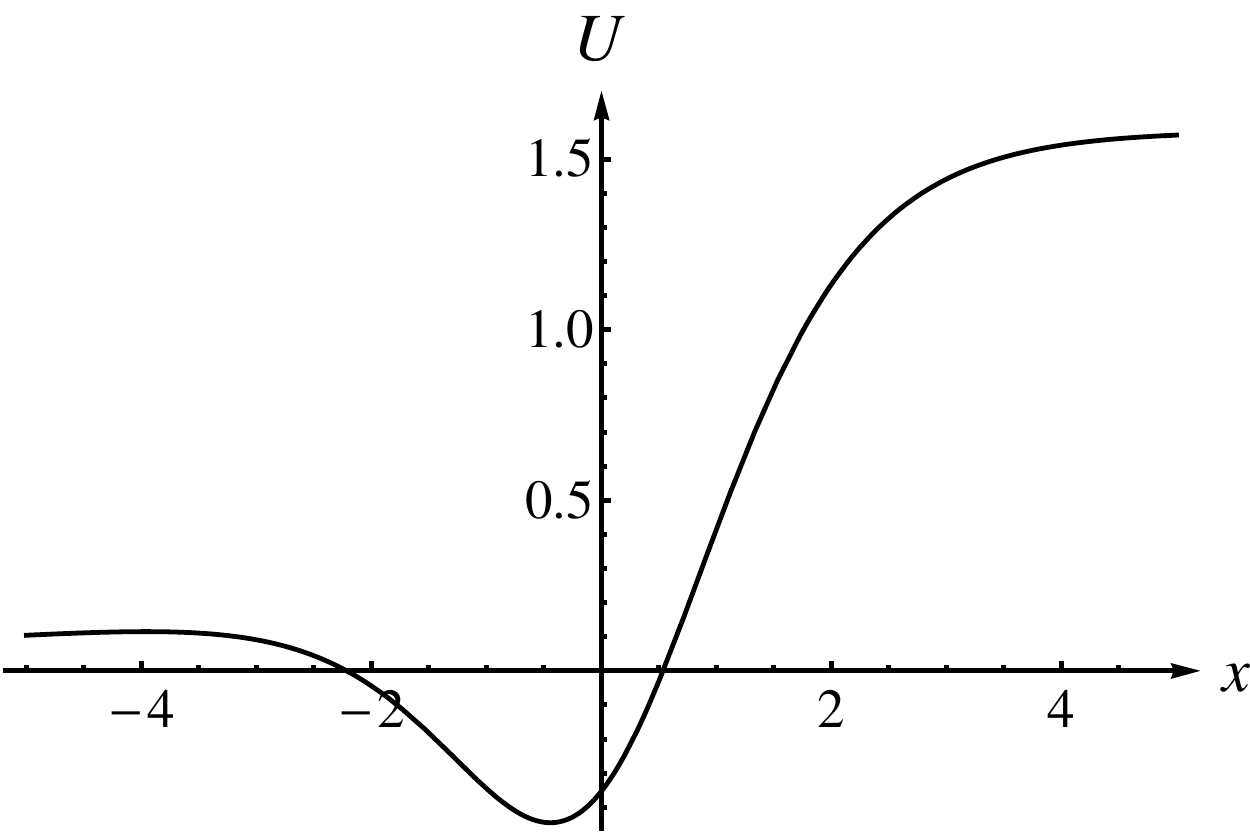}}
\end{minipage}
\caption{{\bf Left panel} --- the kink (\ref{eq:kinkph8c}) that connects the vacua $0$ and $a$.
{\bf Right panel} --- the potential $U(x)$ corresponding to the kink in the left panel.}
\label{fig:kink3}
\end{figure}

This situation corresponds to the potential taking the form
\begin{equation}
V(\varphi)=\lambda^2\varphi^2(\varphi^2-a^2)^2(\varphi^2+b^2),
\label{eq:potphi8b}
\end{equation}
where the constants satisfy $a>0$, $b>0$, $\lambda>0$. This potential has three degenerate
minima: $\overline{\varphi}_1=-a$, $\overline{\varphi}_2=0$, and $\overline{\varphi}_3=a$.
The corresponding topological sectors are $(-a,0)$ and $(0,a)$. Following ref.~\cite{khare},
our choice of parameters is
\[
a=\displaystyle\frac{3}{4},~~~b=1.
\]
Fig.~\ref{fig:potkink2} shows the plot of the resulting potential.
Obviously, the sectors $(-a,0)$ and $(0,a)$ are related by a change of the sign of $\varphi(t,x)$,
hence only one of the two sectors has to be studied. Eq.~(\ref{eq:steomv}) with the potential
(\ref{eq:potphi8b}) results in the following implicit expression for the kink
that connects $\overline{\varphi}_2=0$ and $\overline{\varphi}_3=a$:
\begin{equation}
e^{\mu 
x}=\left(\frac{\sqrt{b^2+a^2}+\sqrt{b^2+\varphi^2}}{\sqrt{b^2+a^2}-\sqrt{b^2+\varphi^2}}\right)\cdot\left(\frac{\sqrt{b^2+\varphi^2}-b}{\sqrt{
b^2+\varphi^2}+b}\right)^{\sqrt{b^2+a^2}/b},
\label{eq:kinkph8c}
\end{equation}
where $\mu=2\sqrt{2}\lambda a^2\sqrt{a^2+b^2}$, with the mass of this kink being
\[
 M_{(0,a)}=\frac{\sqrt{2}}{15}\lambda\left(2(b^2+a^2)^{5/2}-b^3(2b^2+5a^2)\right).
\]
Fig.~\ref{fig:kink3} shows a plot of the kink (\ref{eq:kinkph8c}) in the left panel
and of the corresponding Schr\"odinger potential in the right panel.
Our study of the excitation spectrum of this kink found only the translational mode,
with the numerical result for the eigenvalue $\omega_0^2= 4\times 10^{-8}$.

\subsection{\label{sec:level3C} Two degenerate minima}

\begin{figure}[htb]
\begin{center}
	\includegraphics[scale=0.5]{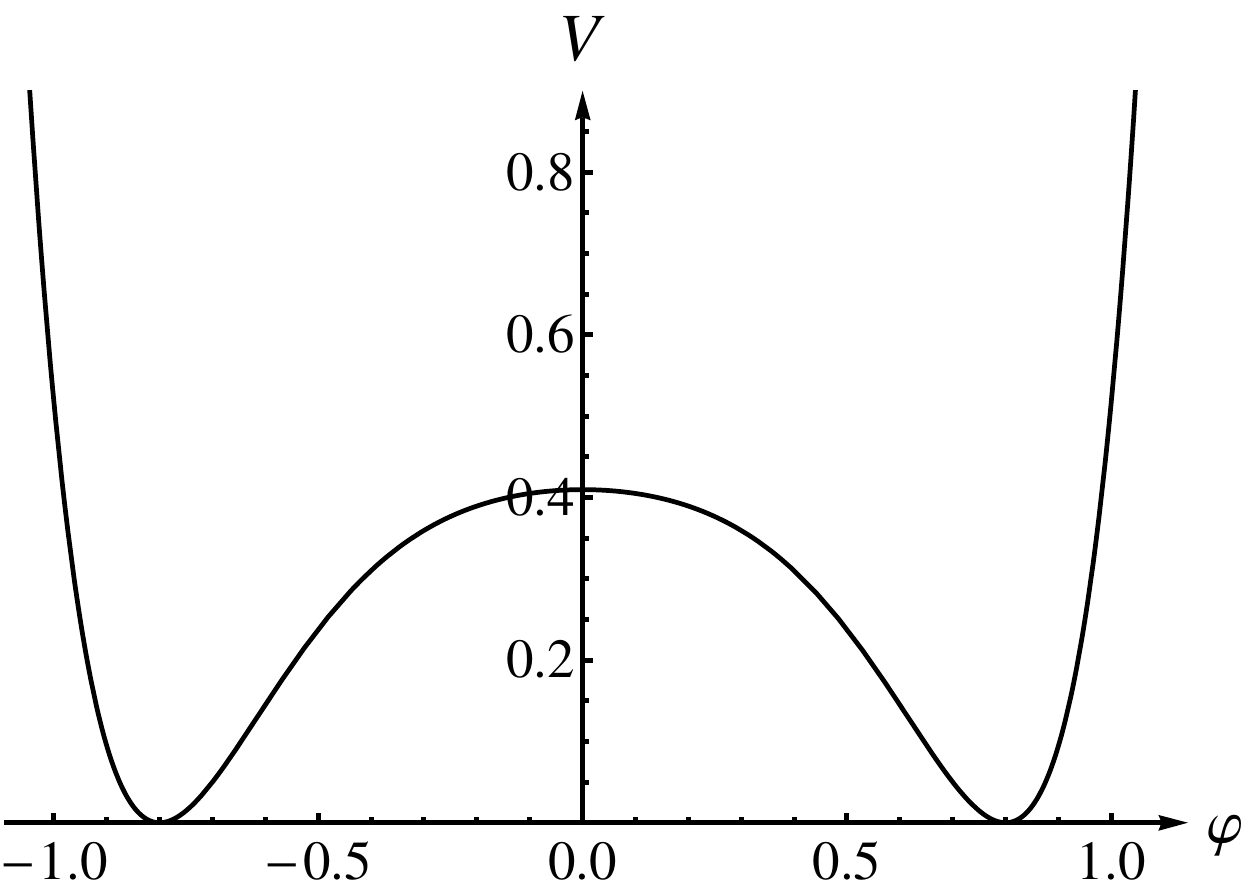}
\end{center}
	\caption{Plot of the potential with two degenerate minima~(\ref{eq:potphi8c}).}
	\label{fig:potkink3}
\end{figure}

The self-interaction potential of the $\varphi^8$ model can in this case be written as
\begin{equation}
V(\varphi)=\lambda^2(\varphi^2-a^2)^2(\varphi^2+b^2)^2,
\label{eq:potphi8c}
\end{equation}
where $a>0$, $b>0$, $\lambda>0$. The two degenerate minima are $\overline{\varphi}_1=-a$ and
$\overline{\varphi}_2=a$, and there is only one kink that connects the points $-a$ and $a$,
and its antikink counterpart. The parameters we use are~\cite{khare}:
\[
a=\displaystyle\frac{4}{5},~~~b=1.
\]
The potential~(\ref{eq:potphi8c}) with these parameters is plotted in Fig.~\ref{fig:potkink3}. 

\begin{figure}[h]
\begin{minipage}[h]{0.49\linewidth}
\center{\includegraphics[width=0.9\linewidth]{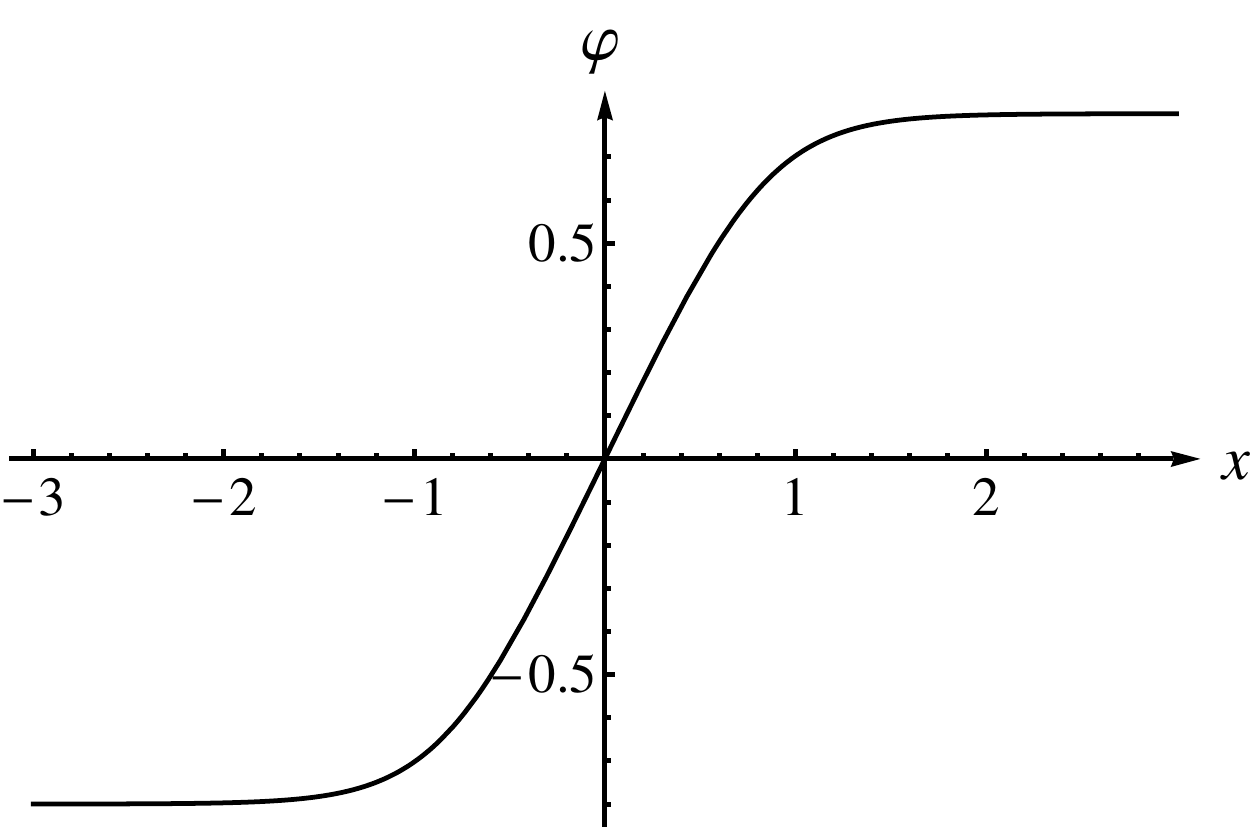}}
\end{minipage}
\hfill
\begin{minipage}[h]{0.49\linewidth}
\center{\includegraphics[width=0.9\linewidth]{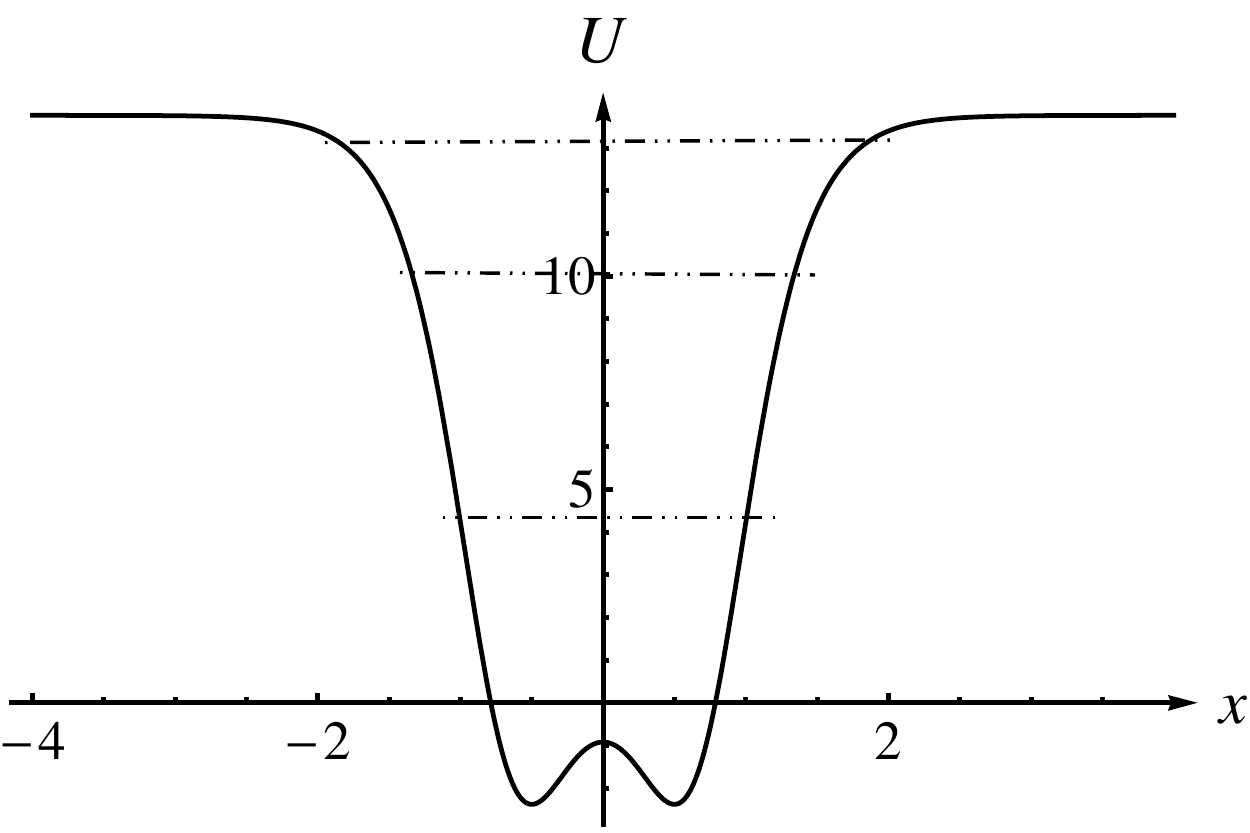}}
\end{minipage}
\caption{{\bf Left panel} --- the kink~(\ref{eq:kinkph8d}) that connects the vacua $-a$ and $a$.
{\bf Right panel} --- the potential $U(x)$ that corresponds to this kink.
The dashed lines show the values of $\omega_1^2$, $\omega_2^2$, and $\omega_3^2$,
ordered from bottom to top.}
\label{fig:kink4}
\end{figure}

The static kink belonging to the topological sector $(-a,a)$ is, as usual, obtained from the equation
of motion~(\ref{eq:steomv}):
\begin{equation}
\mu x=\frac{2a}{b}\tan^{-1}\left(\frac{\varphi}{b}\right)+\log{\left(\frac{a+\varphi}{a-\varphi}\right)},
\label{eq:kinkph8d}
\end{equation}
where $\mu=2\sqrt{2}\lambda a(b^2+a^2)$.
The mass of this kink is
\[
M_{(-a,a)}=\frac{4\sqrt{2}}{15}\lambda a^3(a^2+5b^2).
\]
Fig.~\ref{fig:kink4} shows the kink~(\ref{eq:kinkph8d}) in the left panel, and the corresponding
quantum-mechanical potential that defines small excitations of this kink in the right panel.

The kink of this specific realisation of the $\varphi^8$ model yields the richest set of eigenmodes: besides the translational mode (with the numerical value of $\omega_0^2= 5\cdot 10^{-9}$), we found three vibrational excitations with energies $\omega_1^2= 4.27575$, $\omega_2^2= 10.1893$, and $\omega_3^2= 13.6095$.

A few more words are in order regarding the numerical accuracy. As stated above, the deviation
of the numerical result for $\omega_0^2$ from zero can serve as an estimate of the numerical
error of the energy values. On the other hand, regarding the precision of the
excitation profiles $\psi(x)$, one can, for instance, check that the profiles
corresponding to different eigenvalues (where there are excitations other than the translational
mode) are orthogonal.
The results of this check are given in Table~\ref{tb:table1}.

\begin{table}[h!]
\caption{Eigenvalues and orthogonality checks (where applicable) of the kinks considered in Sec.~\ref{sec:level3}; $\psi_i(x)$ are normalised to unity.}
\begin{center}
\begin{tabular}{|c|c|c|c|c|}
\hline
Kink & $x_l$ & $x_r$ & Eigenvalue & Orthogonality check \\
\hline
\cline{4-4}
\multirow{2}{*}{~(\ref{eq:kinkph8a})~} & \multirow{2}{*}{~$-9$~} & \multirow{2}{*}{~$9$~} &  $\omega_0^2\simeq -2\times 10^{-8}$ & --- \\
\cline{4-5}
           &           &           & $\omega_1^2\simeq 2.70491$ &  $(\psi_0,\psi_1)\simeq -1\times 10^{-7}$      \\
\hline
(\ref{eq:kinkph8b})~&~$-9$~&~$9$~&~$\omega_0^2\simeq -7\times 10^{-8}$~&  --- \\
\hline
~(\ref{eq:kinkph8c})~ &~$-42$~&~$13$~&~$\omega_0^2\simeq 4\times 10^{-8}$~&  --- \\
\hline
\multirow{6}{*}{~(\ref{eq:kinkph8d})~} & \multirow{6}{*}{~$-10$~} & \multirow{6}{*}{~$10$~} & $~\omega_0^2\simeq 5\times 10^{-9}$~ &  --- \\
\cline{4-5}
           &            &            & $\omega_1^2\simeq 4.27575$ & 
           $(\psi_0,\psi_1) \simeq 1\times 10^{-10}$ \\
\cline{4-5}
           &            &            &\multirow{2}{*}{ $\omega_2^2\simeq 10.1893$}  & $(\psi_0,\psi_2) \simeq 4 \times 10^{-8}$ \\
& & & &          $(\psi_1,\psi_2) \simeq 2\times 10^{-9}$ \\
\cline{4-5}
           &            &            & \multirow{3}{*}{$\omega_3^2\simeq 13.6095$}  &~$(\psi_0,\psi_3)\simeq -4\times 10^{-8}$~ \\
           &            &            &      & $(\psi_1,\psi_3)\simeq 2\times 10^{-5}$ \\
           &            &            &     & $(\psi_2,\psi_3)\simeq 2\times 10^{-6}$ \\
\hline
\end{tabular}
\end{center}
\label{tb:table1}
\end{table}

\section{\label{sec:level4} Kink-kink collisions}

The existence of a vibrational excitation of a kink means that, in certain conditions,
kinetic energy of the moving kink can be transferred to and stored in the form
of small oscillations of the kink's profile. This is known to lead to interesting phenomena in kink-kink collisions~\cite{aek01,krusch01,dorey01,GaKuPRE,campbell,peyrard,GaKuLi}.

As an illustrative example, we performed a numerical calculation of the kink-antikink scattering
in one of the topological sectors of the variant of the $\varphi^8$ theory with four degenerate minima,
i.e., with the field self-interaction given by Eq.~(\ref{eq:potphi8}).
We studied a configuration constructed from the kink (\ref{eq:kinkph8a})
and the corresponding antikink (\ref{eq:kinkph8ak}). Namely, we set
the initial field profile to
\begin{equation}
\varphi(t,x)=\varphi_{(-a,a)}\left(\frac{x+x_0-v_{\scriptsize \mbox{in}}t}{\sqrt{1-v^2_{\scriptsize 
\mbox{in}}}}\right)+\varphi_{(a,-a)}\left(\frac{x-x_0+v_{\scriptsize \mbox{in}}t}{\sqrt{1-v^2_{\scriptsize \mbox{in}}}}\right)-a,
\label{eq:phiin}
\end{equation}
where $\varphi_{(-a,a)}(x)$ and $\varphi_{(a,-a)}(x)$ are the static kink and the static antikink.
This setup corresponds to the kink and the antikink, separated at $t=0$ by $2x_0$, and moving towards each
other with the velocities $\pm v_\mathrm{in}$ in the laboratory frame.
As the separation between the kinks increases, the overlap between them becomes exponentially small and they stop being affected by each other --- this happens in the asymptotic regime of the collision. In practice,
the initial separation $2x_0$ has to be much larger than the typical
width of the kink; in our calculation, we use $2x_0=25$.

We solved the equation of motion using the standard explicit finite difference scheme,
\[
\varphi_{tt}=\frac{\varphi_j^{k+1}-2\varphi_j^k+\varphi_j^{k-1}}{\tau^2},~~\varphi_{xx}=\frac{\varphi_{j+1}^{k}-2\varphi_j^k+\varphi_{j-1}^{k}}{h^2},
\]
where $\tau$ and $h$ are, respectively, the time and space grid spacings, and $(k, j)$ number the corresponding coordinates of the grid points, $(t_k,x_j)$.
The initial conditions follow from Eq.~(\ref{eq:phiin}), and the presented results correspond to $\tau=0.002$ and $h=0.01$.
The infinite space domain was truncated, so that $-l\le x \le l$, with $l=100$, whereas the time varied in the range $0\le t\le 900$.
As dictated by the dependence domains, the calculation
started from a much larger interval of the $x$ axis at $t=0$ (namely, $-4650\le x\le 4650$).
To check our numerical results, we tested the conservation of energy
as the time evolution progressed, taking into account energy
flux through the endpoints of the space interval:
\[
\int\limits_{-l}^{l}\left[\frac{1}{2} \left( \frac{\partial\varphi}{\partial t} \right)^2+\frac{1}{2} \left( \frac{\partial\varphi}{\partial x} \right) ^2
+V(\varphi)\right]dx-\int\limits_{0}^{t_c} \frac{\partial\varphi}{\partial t}\frac{\partial\varphi}{\partial x} \biggl|_{-l}^l dt=\frac{2M}{\sqrt{1-v_{\scriptsize \mbox{in}}^2}},
\]
where $t_c$ is the current moment of time (the integrand in the first integral is evaluated at $t=t_c$).

\begin{figure}[h]
\begin{minipage}[h]{0.49\linewidth}
\center{\includegraphics[width=0.9\linewidth]{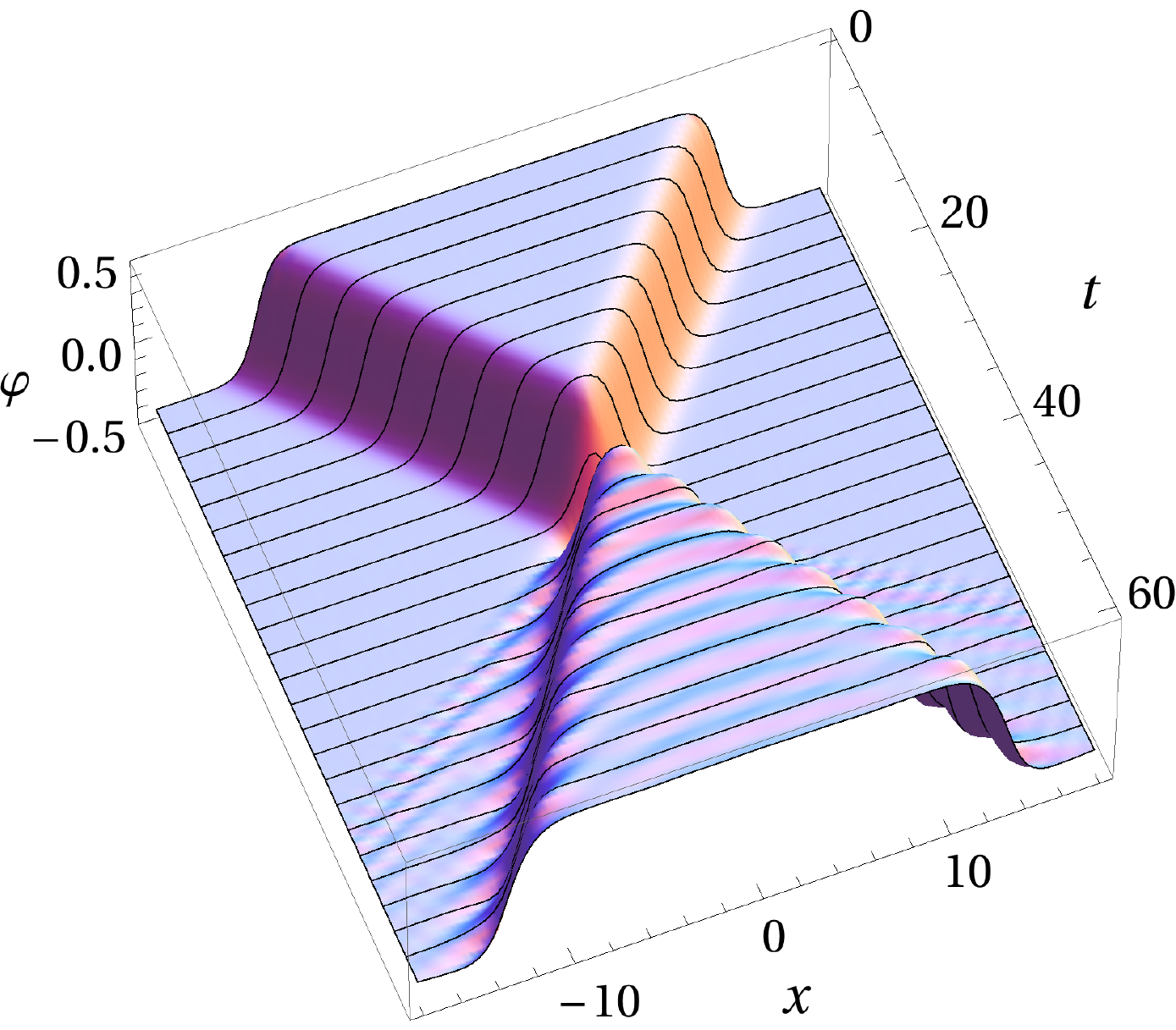}}
\end{minipage}
\hfill
\begin{minipage}[h]{0.49\linewidth}
\center{\includegraphics[width=0.9\linewidth]{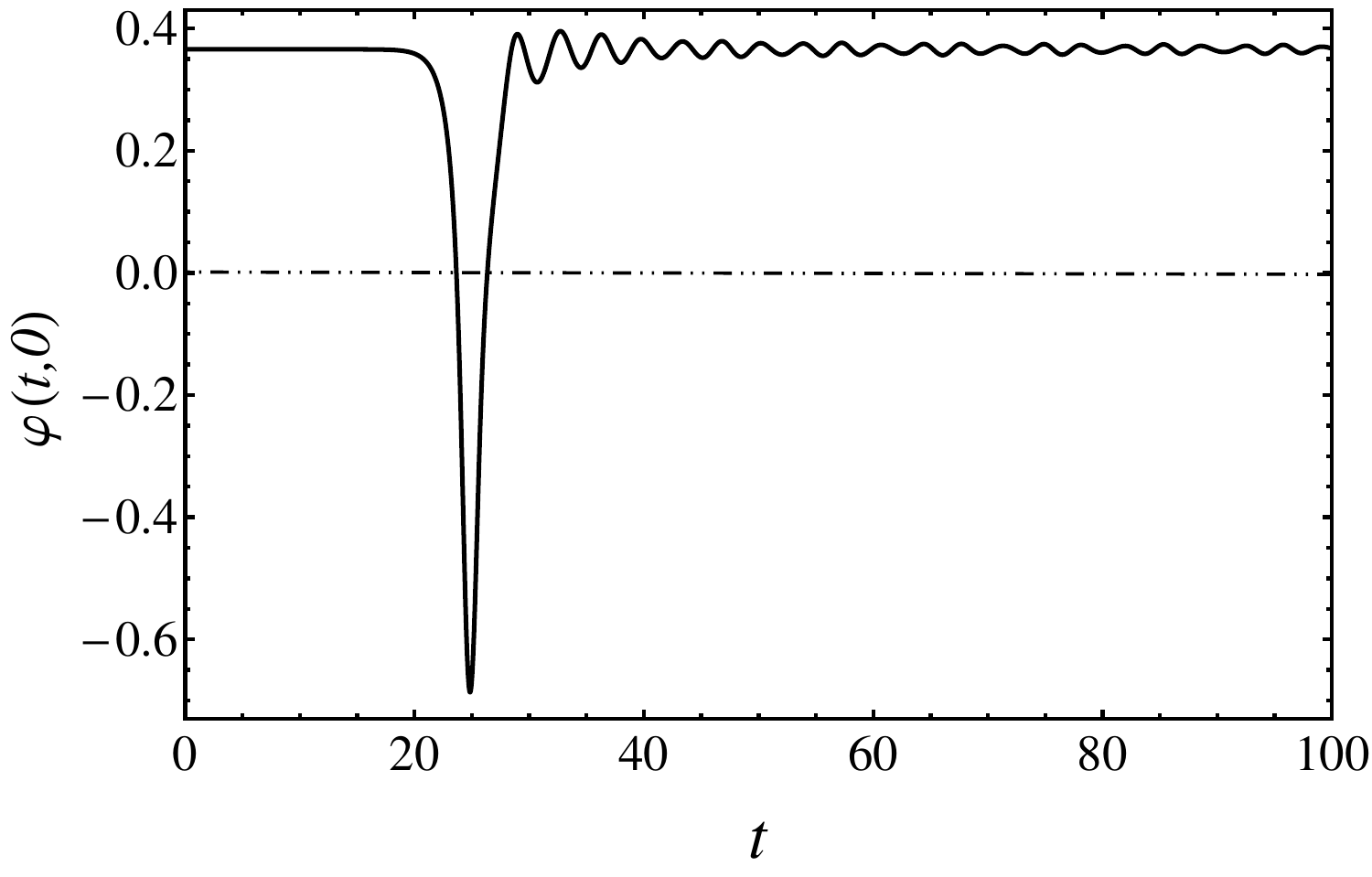}}
\end{minipage}
\caption{(Inelastic) kink-kink scattering at $v_{\scriptsize \mbox{in}}= 0.5000$. {\bf Left panel} --- space-time evolution of the system. {\bf Right panel} --- $\varphi(t,0)$, the time profile of the field at $x=0$.}
\label{fig:razlet5000}
\end{figure}

Our numerical simulations showed several different scattering regimes
can realise, depending on the initial collision speed. We found out that
there is a critical speed $v_{\scriptsize \mbox{cr}}\simeq 0.3160$
such that at $v_{\scriptsize \mbox{in}}\ge v_{\scriptsize \mbox{cr}}$
the kinks bounce off each other and escape, as illustrated in Fig.~\ref{fig:razlet5000}. The kink-antikink scattering
is only approximately elastic, i.e., the asymptotic escape velocity 
$v_{\scriptsize \mbox{f}}<v_{\scriptsize \mbox{in}}$. This feature can be seen
in Fig.~\ref{fig:razlet5000} as well; the typical values are, e.g., $v_\mathrm{f}=0.23$
at $v_\mathrm{in}=0.40$, $v_\mathrm{f}=0.36$ at $v_\mathrm{in}=0.50$, and
$v_\mathrm{f}=0.47$ at $v_\mathrm{in}=0.60$.

\begin{figure}[h]
\begin{minipage}[h]{0.49\linewidth}
\center{\includegraphics[width=0.9\linewidth]{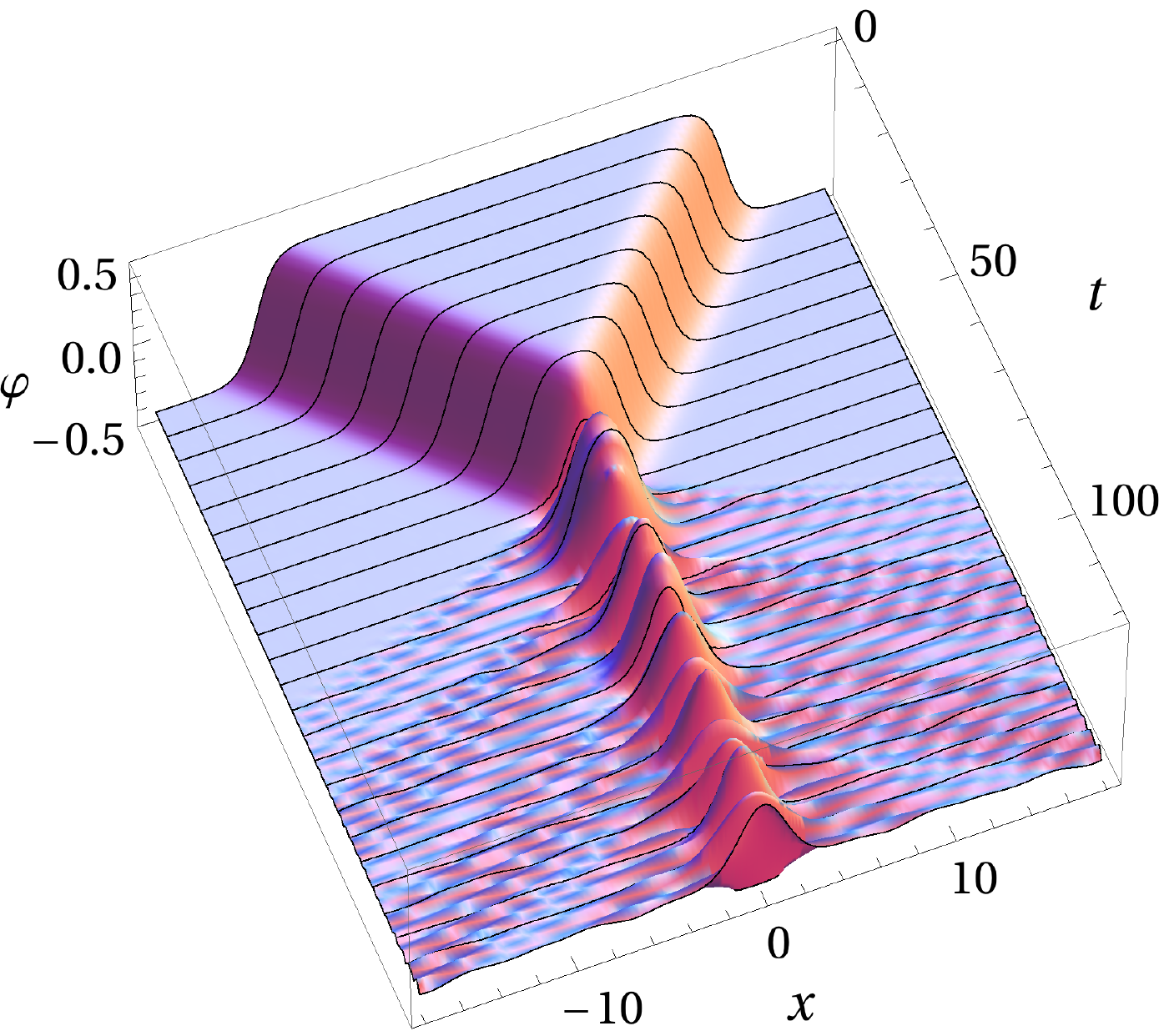}}
\end{minipage}
\hfill
\begin{minipage}[h]{0.49\linewidth}
\center{\includegraphics[width=0.9\linewidth]{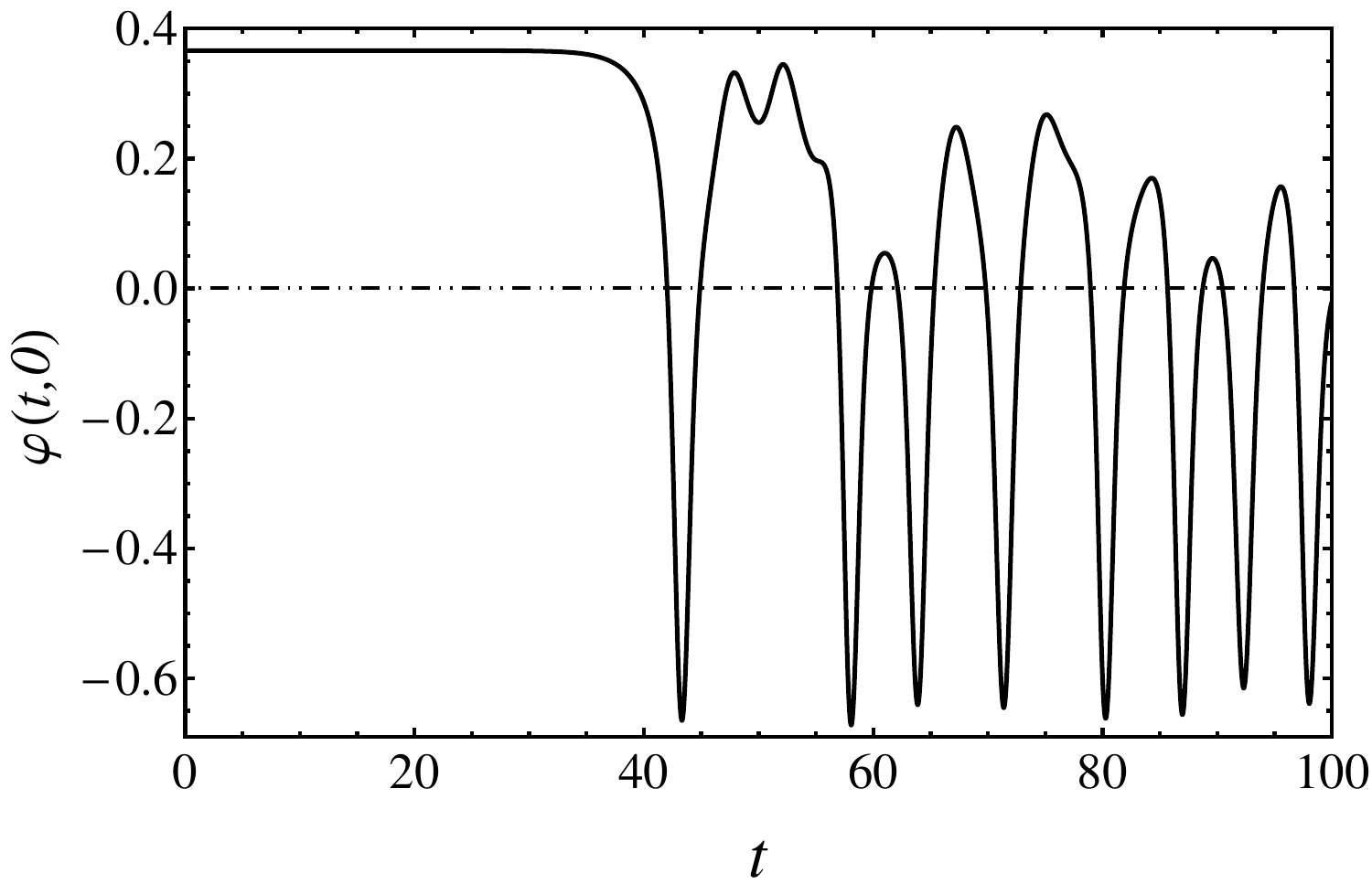}}
\end{minipage}
\caption{Bion formation at $v_{\scriptsize \mbox{in}}= 0.2740$. {\bf Left panel} --- space-time evolution. {\bf Right panel} --- the profile of $\varphi(t,0)$.}
\label{fig:bion2740}
\end{figure}

\begin{figure}[h]
\begin{minipage}[h]{0.49\linewidth}
\center{\includegraphics[width=0.9\linewidth]{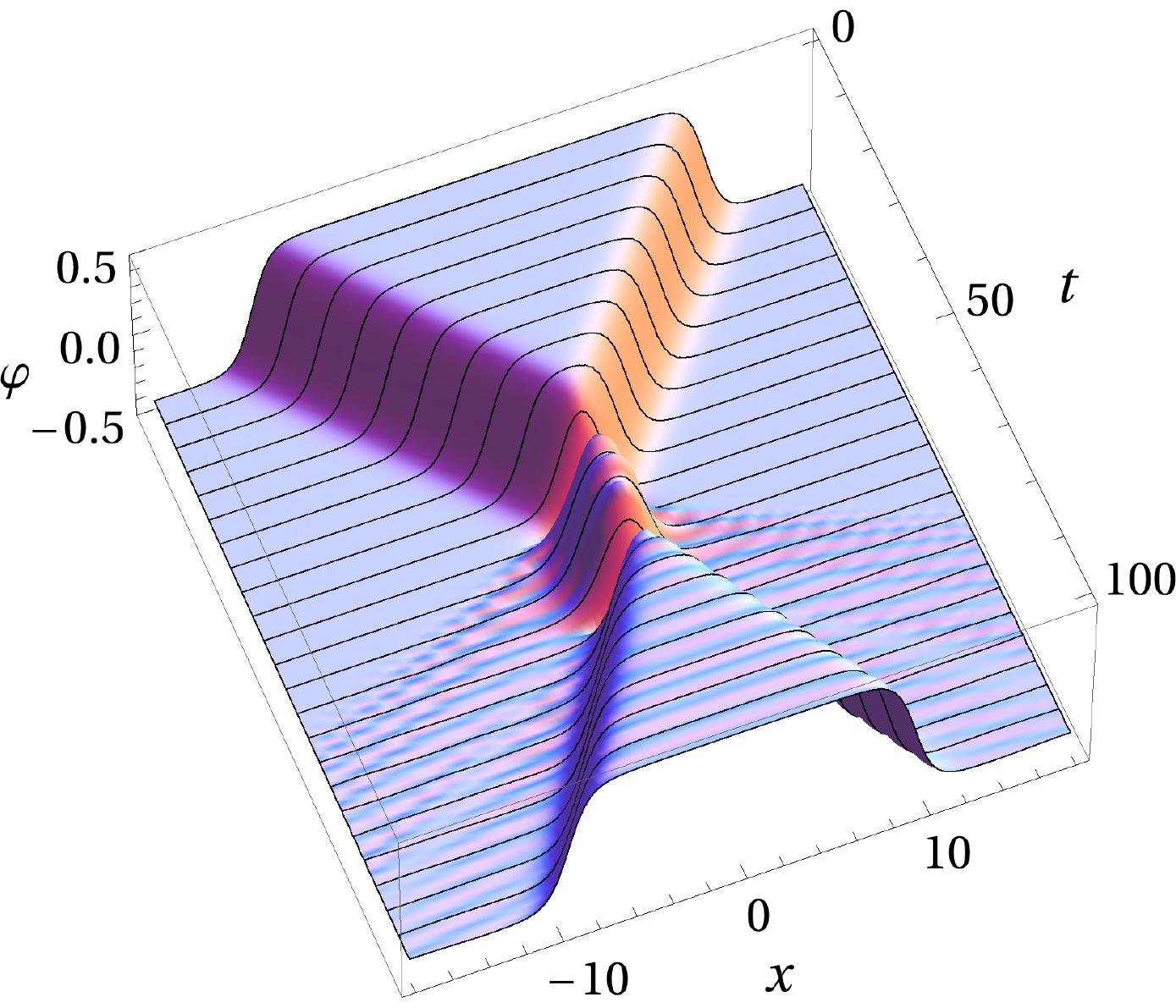}}
\end{minipage}
\hfill
\begin{minipage}[h]{0.49\linewidth}
\center{\includegraphics[width=0.9\linewidth]{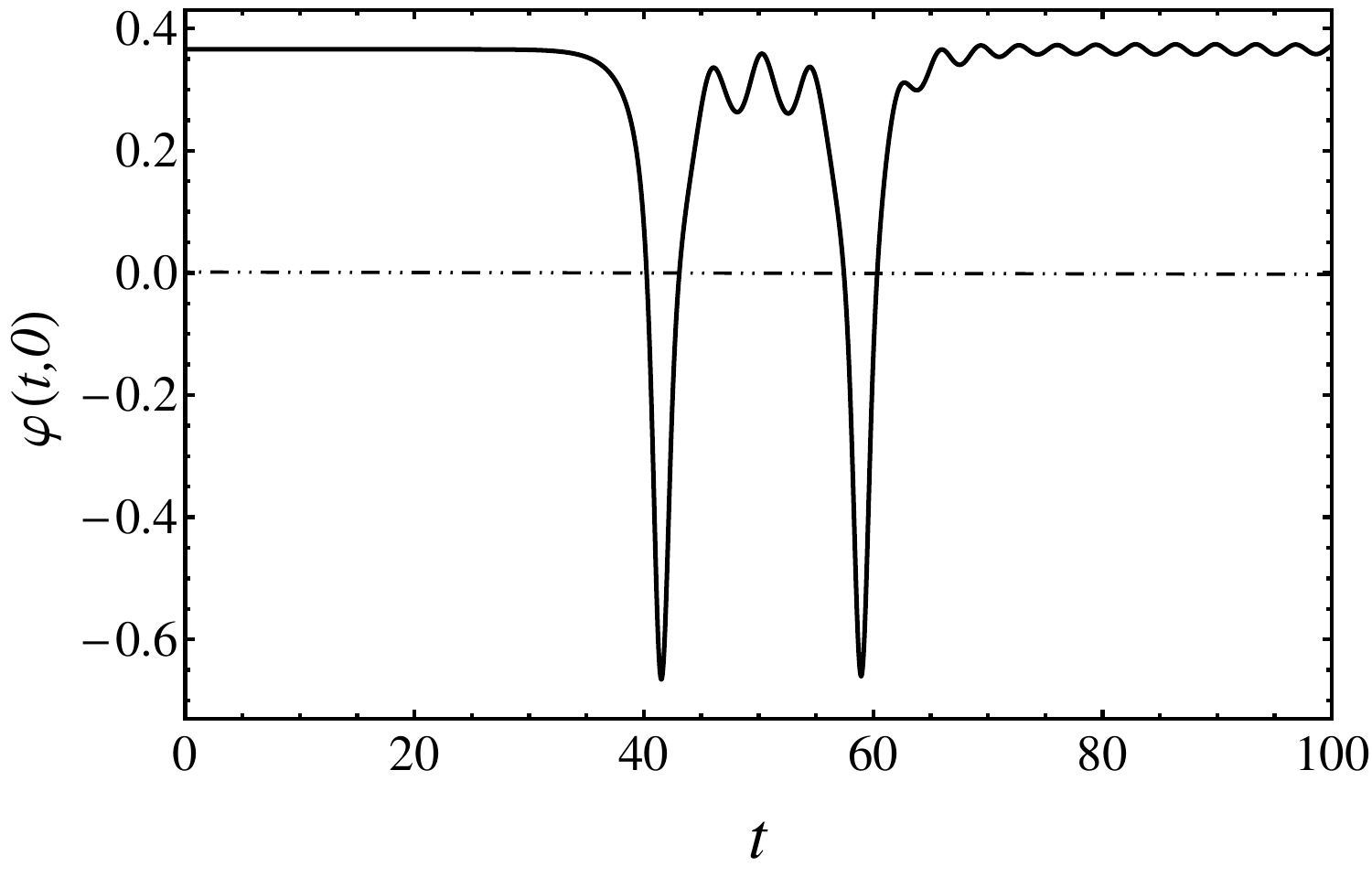}}
\end{minipage}
\caption{Two-bounce at $v_{\scriptsize \mbox{in}}= 0.2868$. {\bf Left panel} --- space-time evolution. {\bf Right panel} --- the profile of $\varphi(t,0)$.}
\label{fig:window2_2868}
\end{figure}

\begin{figure}[h]
\begin{minipage}[h]{0.49\linewidth}
\center{\includegraphics[width=0.9\linewidth]{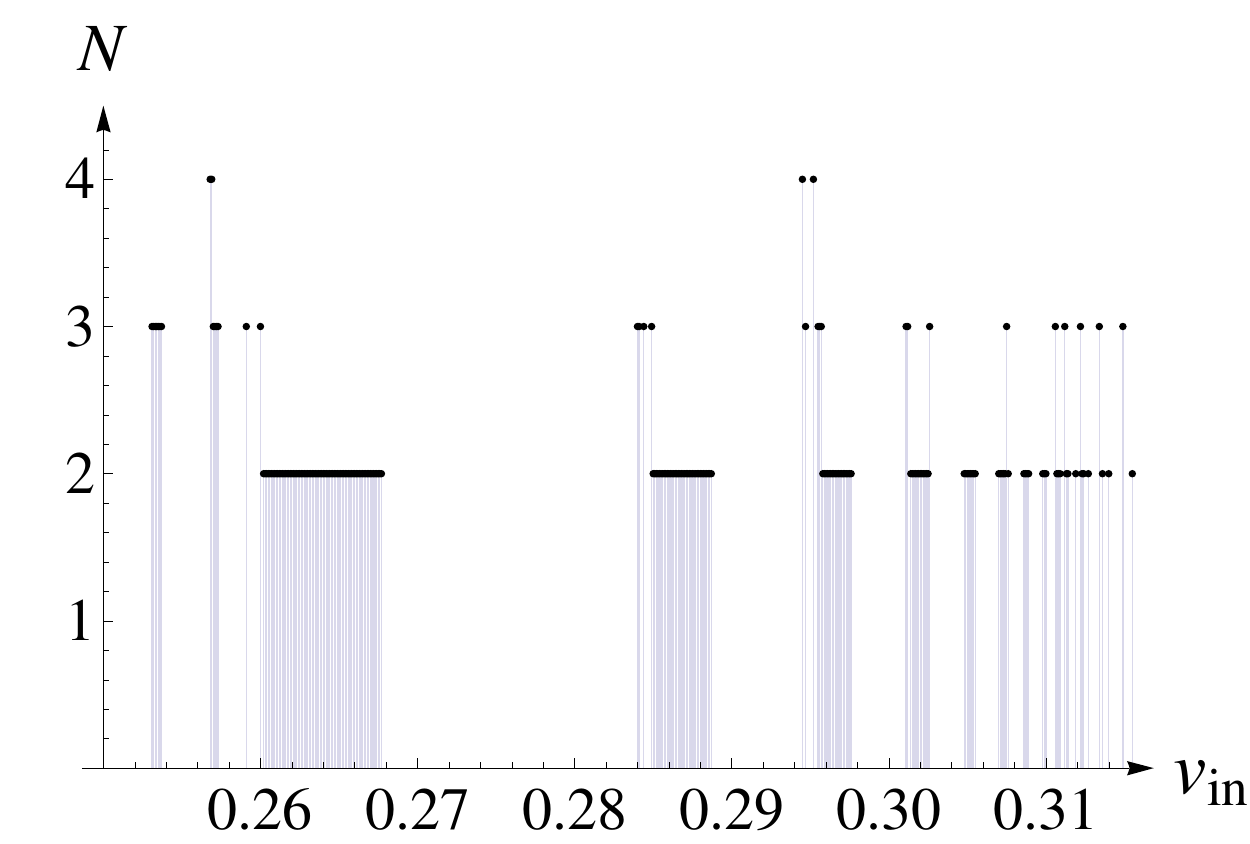}}
\end{minipage}
\hfill
\begin{minipage}[h]{0.49\linewidth}
\center{\includegraphics[width=0.9\linewidth]{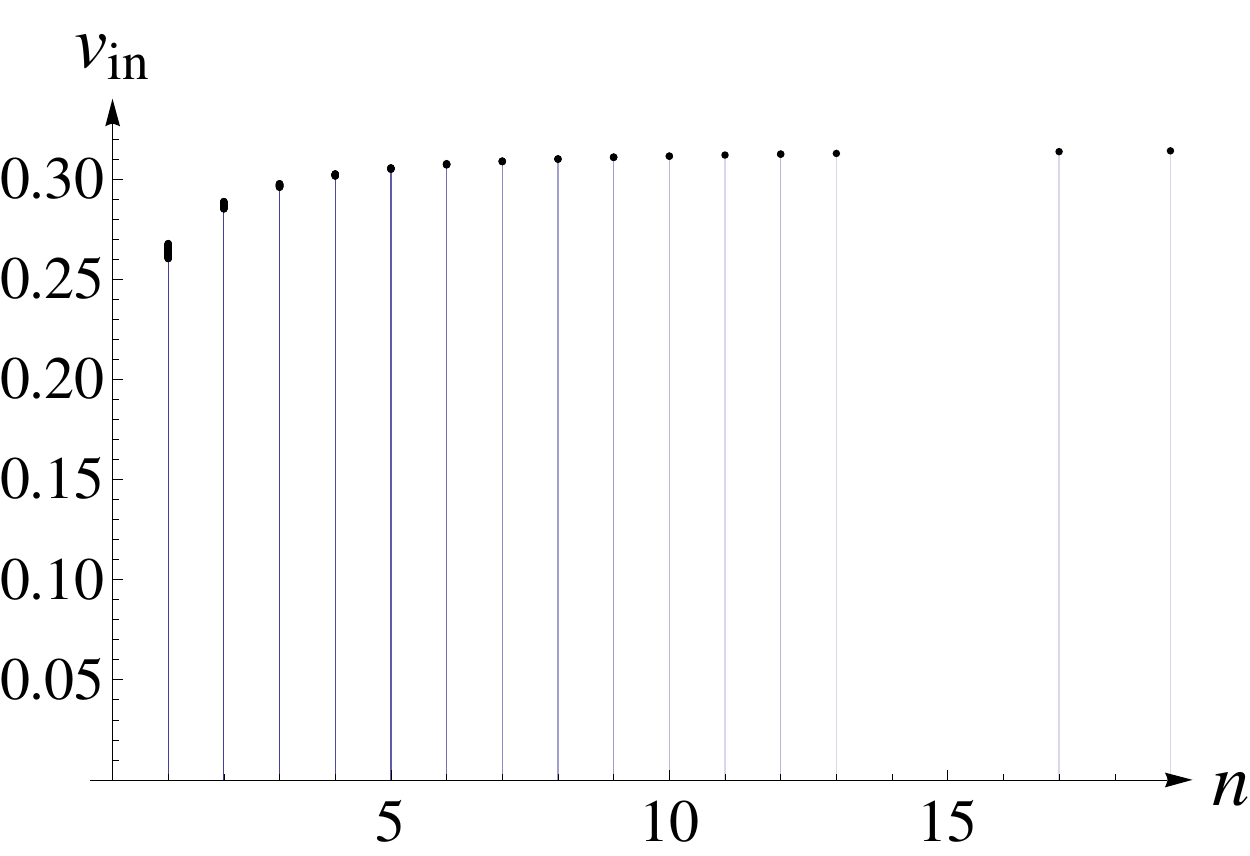}}
\end{minipage}
\caption{{\bf Left panel} --- Quasi-fractal structure
of the escape windows; here $N$ is the number of collisions before escape. {\bf Right panel} ---
locations of two-bounce escape windows versus $n$,
the number of small oscillations (see text for
the definition).}
\label{fig:escwindow}
\end{figure}

\begin{figure}[h]
\begin{center}
\includegraphics[scale=0.6]{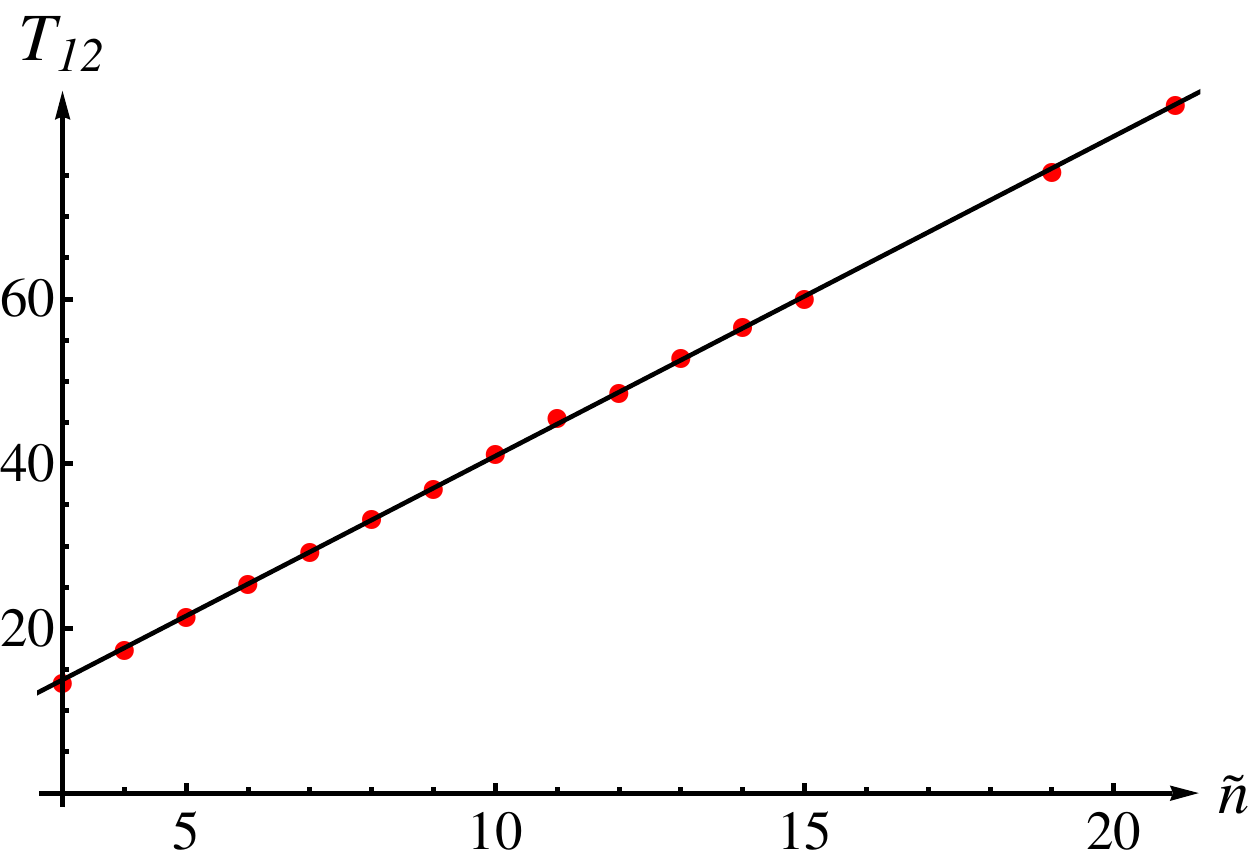}
\end{center}
\caption{Time between the first two collisions $T_{12}$ as a function of $\tilde n$
for two-bounce escape windows. Dots show results of our numerical
calculations, the solid line is the corresponding linear fit of Eq.~(\ref{eq:rescond}).}
\label{fig:t12n}
\end{figure}

As opposed to the previous regime, collision velocities $v_{\scriptsize \mbox{in}}<v_{\scriptsize \mbox{cr}}$ result in the formation of a long-lived bound state of the two kinks, the bion, see Fig.~\ref{fig:bion2740} (as observed long ago, for instance, in the $\varphi^4$ model~\cite{Kudbion}). This, however, does not hold for all
$v_{\scriptsize \mbox{in}}<v_{\scriptsize \mbox{cr}}$; there are intervals of the collision velocities --- the escape windows --- where kinks still escape to the spatial infinity albeit only after having collided two, three, or more times. Similar
processes have been observed in other field models, see, e.g.,~\cite{aek01} for review.
This phenomenon has been explained by resonant energy exchange between
the translational mode and a localized excitation mode of the kink.
The value of the corresponding resonance frequency $\omega_R$ in some models coincides with the kink
vibrational mode frequency $\omega_1$,
whereas in other field models $\omega_R$ could significantly deviate from
$\omega_1$~\cite{GaKuPRE,campbell}, or even
belong to the continuous part of the excitation spectrum~\cite{peyrard}.
During the first collision, the kinetic energy of the kinks can partly
be transferred to the mode $\omega_R$. After that, the kinks bounce
but cannot escape, and hence they stop at some point and then collide again. However, the second collision may result in transfer of the
energy of the mode $\omega_R$ back to the kinetic energy.
This can happen if the time $T_{12}$ between the collisions
and the frequency $\omega_R$ are in a certain
resonance relation (see below), and, as a result, the kinks can
escape after the second collision.
An example of such a two-bounce
is shown in Fig.~\ref{fig:window2_2868}.
Note that this regime
is also inelastic --- the kinks lose a part
of their initial kinetic energy; this can be seen in the figures as well.
We found around twenty two-bounce escape windows,
as well as a number of three- and four-bounce escape windows (where the kinks escape to
infinity after three and four collisions). Fig.~\ref{fig:escwindow} shows the positions
of these $\varphi^8$ model escape windows and
demonstrates that they exhibit features similar to observed, e.g., in the
$\varphi^4$ model~\cite{aek01}, namely, the quasi-fractal structure
and the concentration of resonances near the critical velocity $v_\mathrm{cr}$.
We also analysed the positions of the two-bounce escape windows, using phenomenological relations that
have been applied to resonance kink-kink collisions in other models, cf.\ ref.~\cite{aek01}.
The main results of this analysis are shown in Fig.~\ref{fig:t12n} and in Table~\ref{tb:table2}.

The first of the relations in question connect the resonance frequency $\omega_R$
with the number $n$ of small oscillations of the field at $x=0$ between the two collisions
(so that, e.g., the two-bounce shown in Fig.~\ref{fig:window2_2868} corresponds to $n=2$):
\begin{equation}
  \omega_R T_{12}=2\pi(n+2)+\delta=2\pi\tilde n+\delta,
  \label{eq:rescond}
\end{equation}
where $\delta$ is a constant phase shift. Note that $n$ can also be viewed as the
number of the respective escape window in the sequence shown in Fig.~\ref{fig:escwindow}
(implying that the windows with numbers $14$, $15$, $16$, and $18$ are missing --- we have
not observed them in our numerical simulations).
The shift of $n$ by two in Eq.~(\ref{eq:rescond}) is dictated by our choice that $\delta$ has to be
between $0$ and $2\pi$. Fig.~\ref{fig:t12n} shows $T_{12}$ as a function of $\tilde n$. As one can
see from this figure, the linear dependence of Eq.~(\ref{eq:rescond}) fits nicely the numerical results,
with the values of the resonance parameters resulting from the fit being $\omega_R = 1.618$ and
$\delta = 3.403$. This value of $\omega_R$ is slightly less than the frequency of the kink's
vibrational excitation, $\omega_1=1.644$. This can be attributed to the interaction between the two
kinks: it continuously distorts the excitation spectrum of the kink-kink system
during the collision, which (inter alia) can lead to a deviation of $\omega_R$ from
$\omega_1$~\cite{GaKuPRE}.
As mentioned above, the collective excitation spectrum of the kink-kink system can have a vibrational
mode even when either of the solitary kinks does not.
This feature gives rise to resonance phenomena such as escape windows and quasi-resonances
in models such as $\varphi^6$ where the kinks only have the translational excitation~\cite{dorey01}.

Table~\ref{tb:table2} collects information we have obtained on two-bounce escape windows.
It lists the initial collision velocities $v_{\scriptsize \mbox{in}}$ (taken at the midpoint
of the corresponding window), the final escape velocities $v_{\scriptsize \mbox{f}}$, and the
values of $T_{12}$. We also applied the phenomenological analysis of ref.~\cite{peyrard}, calculating
\begin{equation}
\beta = T_{12}\sqrt{v_{\scriptsize \mbox{cr}}^2-v_{\scriptsize \mbox{in}}^2}
\end{equation}
for each of the escape windows, with the mean value being $\bar{\beta}\simeq 2.569$.
The latter quantity was used to predict the locations of escape windows:
\begin{equation}
v_{\scriptsize \mbox{in}}^\mathrm{theor}(n)=v_{\scriptsize \mbox{cr}}-\frac{\bar{\beta}^2\omega_R^2}{2 v_{\scriptsize \mbox{cr}}(2\pi\tilde n + \delta)^2},
\label{eq:vintheor}
\end{equation}
where $v_{\scriptsize \mbox{in}}^\mathrm{theor}$ is the predicted escape window initial velocity,
assuming small deviation of the latter from the critical velocity,
$v_{\scriptsize \mbox{in}} \simeq v_{\scriptsize \mbox{cr}}$.
Table~\ref{tb:table2} shows that Eq.~(\ref{eq:vintheor}) does not provide accurate
predictions for $v_\mathrm{in}$ --- theoretical values at large values of $n$ can differ
from the corresponding calculated values by an amount larger than the distance between two
adjacent escape windows. This is at least partly due to poor accuracy of the determination of
$\bar{\beta}$, which has a scatter of about $15\%$. We have thus not been able to use
Eq.~(\ref{eq:vintheor})
to pinpoint the location of the escape windows with $n=14, 15, 16, 18$ that had not been observed
directly in the numerical calculations.

\begin{table}[h!]
\caption{Two-bounce escape windows. The values of $v_{\scriptsize \mbox{in}}$
are given at the midpoint of the corresponding escape window.}
\begin{center}
\begin{tabular}{|c|c|c|c|c|c|}
\hline
~$n$~ & ~$v_{\scriptsize \mbox{in}}$~ & ~$v_{\scriptsize \mbox{f}}$~ & ~$T_{12}$~ & ~$\beta$~ & ~$v_{\scriptsize \mbox{in}}^\mathrm{theor}$~ \\
\hline
~$1$~ & ~$0.2639$~ & ~$0.158$~ & ~$13.446$~ &$2.337$~&  ~$0.2608$~ \\
\hline
$2$ & $0.2868$ & $0.207$ & $17.422$ &  $2.311$&  $0.2824$ \\
\hline
$3$ & $0.2967$ & $0.229$ & $21.424$ &  $2.330$&  $0.2934$ \\
\hline
$4$ & $0.3019$ & $0.239$ & $25.386$ &  $2.370$&  $0.2998$ \\
\hline
$5$ & $0.3051$ & $0.237$ & $29.256$ &  $2.407$&  $0.3038$ \\
\hline
$6$ & $0.3073$ & $0.247$ & $33.280$ &  $2.451$&  $0.3065$ \\
\hline
$7$ & $0.3087$ & $0.214$ & $36.940$ &  $2.495$&  $0.3084$ \\
\hline
$8$ & $0.3099$ & $0.253$ & $41.180$ &  $2.545$&  $0.3098$ \\
\hline
$9$ & $0.3108$ & $0.079$ & $45.528$ &  $2.599$&  $0.3108$ \\
\hline
$10$ & $0.3113$ & $0.222$ & $48.678$ &  $2.643$&  $0.3116$ \\
\hline
$11$ & $0.3119$ & $0.253$ & $52.892$ &  $2.684$&  $0.3122$ \\
\hline
$12$ & $0.3123$ & $0.249$ & $56.624$ &  $2.730$&  $0.3127$ \\
\hline
$13$ & $0.3127$ & $0.124$ & $60.058$ &  $2.736$&  $0.3131$ \\
\hline
$17$ & $0.3136$ & $0.041$ & $75.466$ &  $2.934$&  $0.3142$ \\
\hline
$19$ & $0.3140$ & $0.221$ & $83.542$ &  $2.965$&  $0.3145$ \\

\hline
\end{tabular}
\end{center}
\label{tb:table2}
\end{table}

\begin{figure}[h]
\begin{minipage}[h]{0.49\linewidth}
\center{\includegraphics[width=0.9\linewidth]{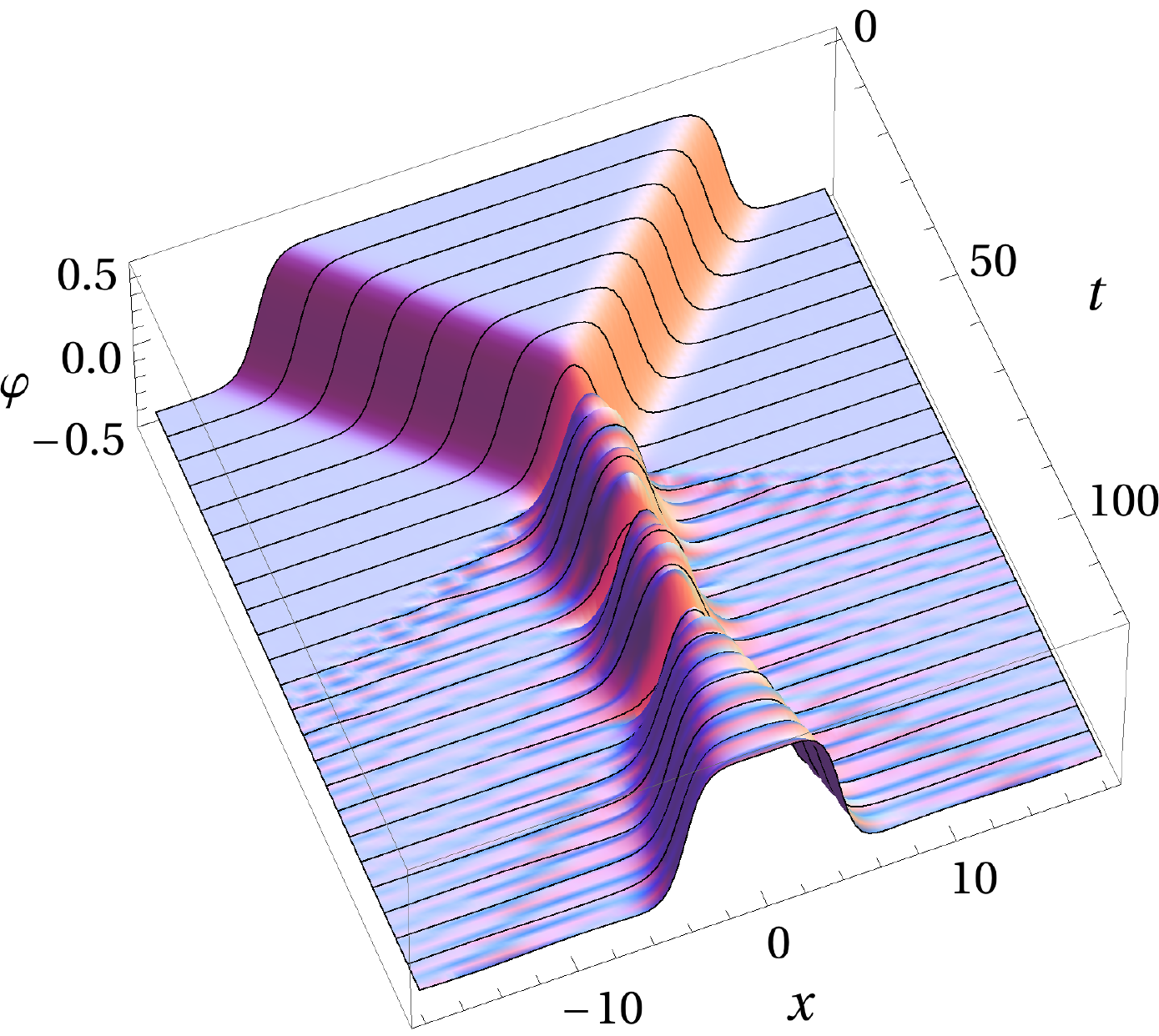}}
\end{minipage}
\hfill
\begin{minipage}[h]{0.49\linewidth}
\center{\includegraphics[width=0.9\linewidth]{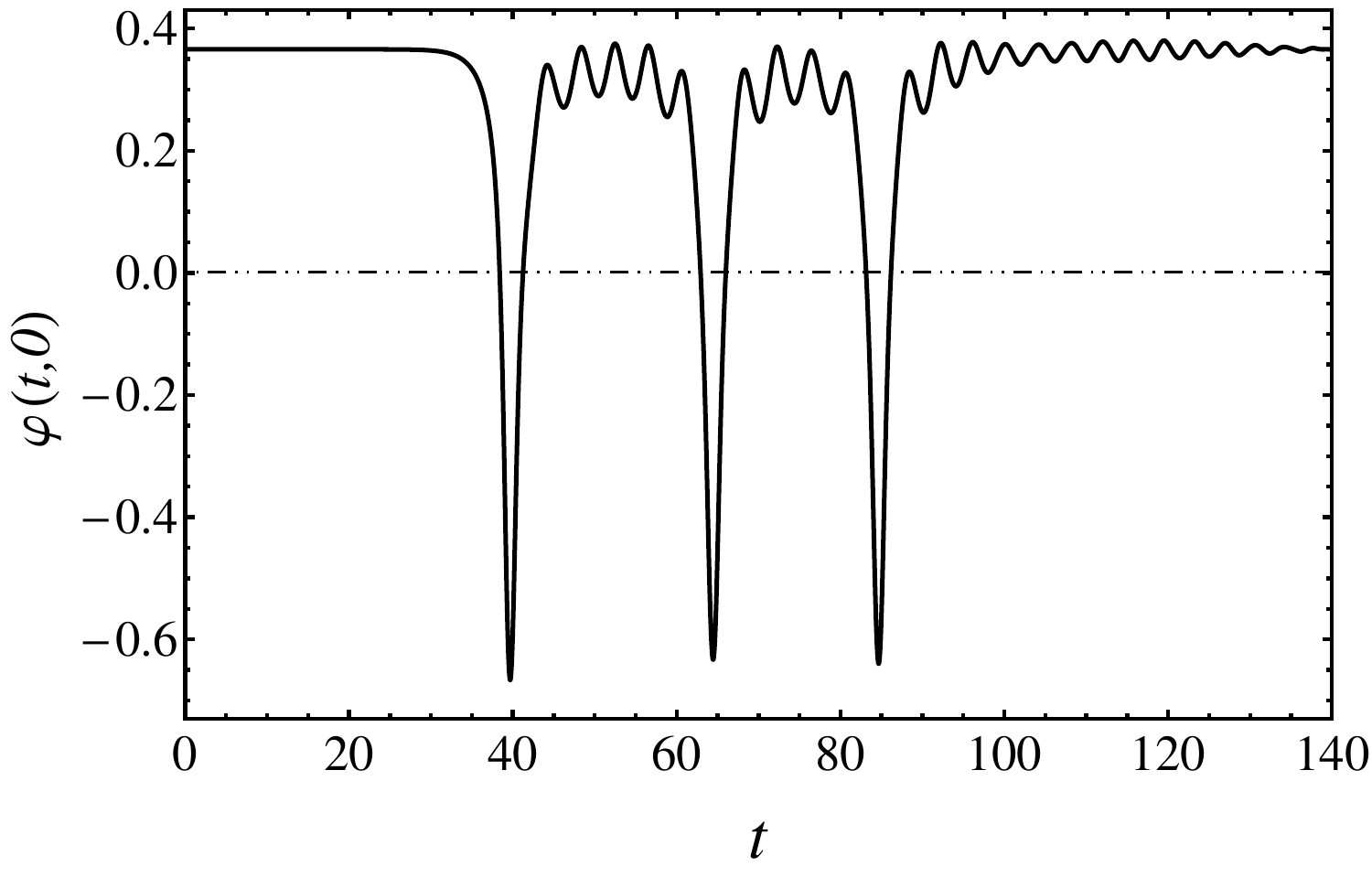}}
\end{minipage}
\caption{Three-bounce at $v_{\scriptsize \mbox{in}}= 0.3012$. {\bf Left panel} --- space-time evolution. {\bf Right panel} --- the profile of $\varphi(t,0)$.}
\label{fig:window3_3012}
\end{figure}

\begin{figure}[h]
\begin{minipage}[h]{0.49\linewidth}
\center{\includegraphics[width=0.9\linewidth]{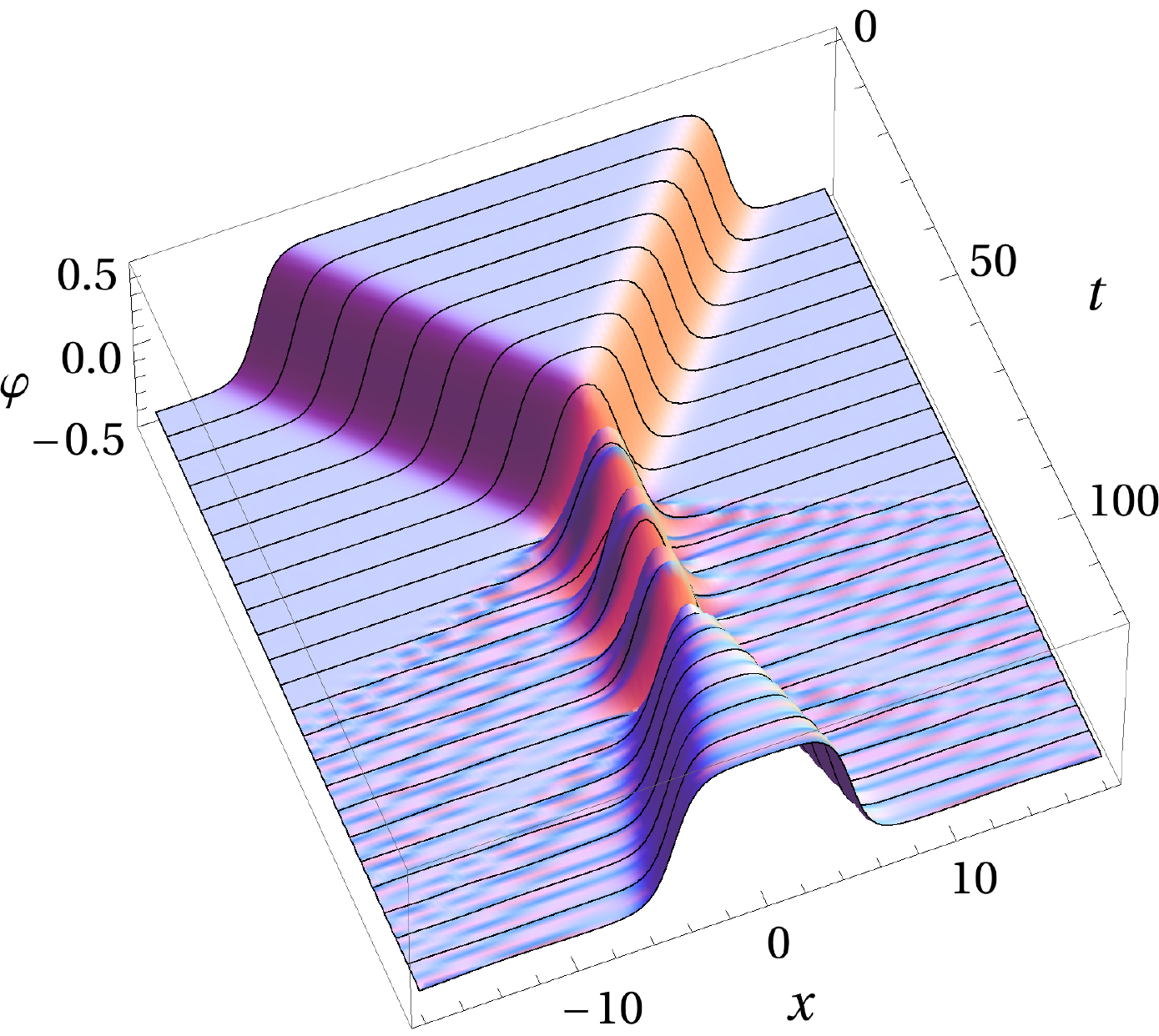}}
\end{minipage}
\hfill
\begin{minipage}[h]{0.49\linewidth}
\center{\includegraphics[width=0.9\linewidth]{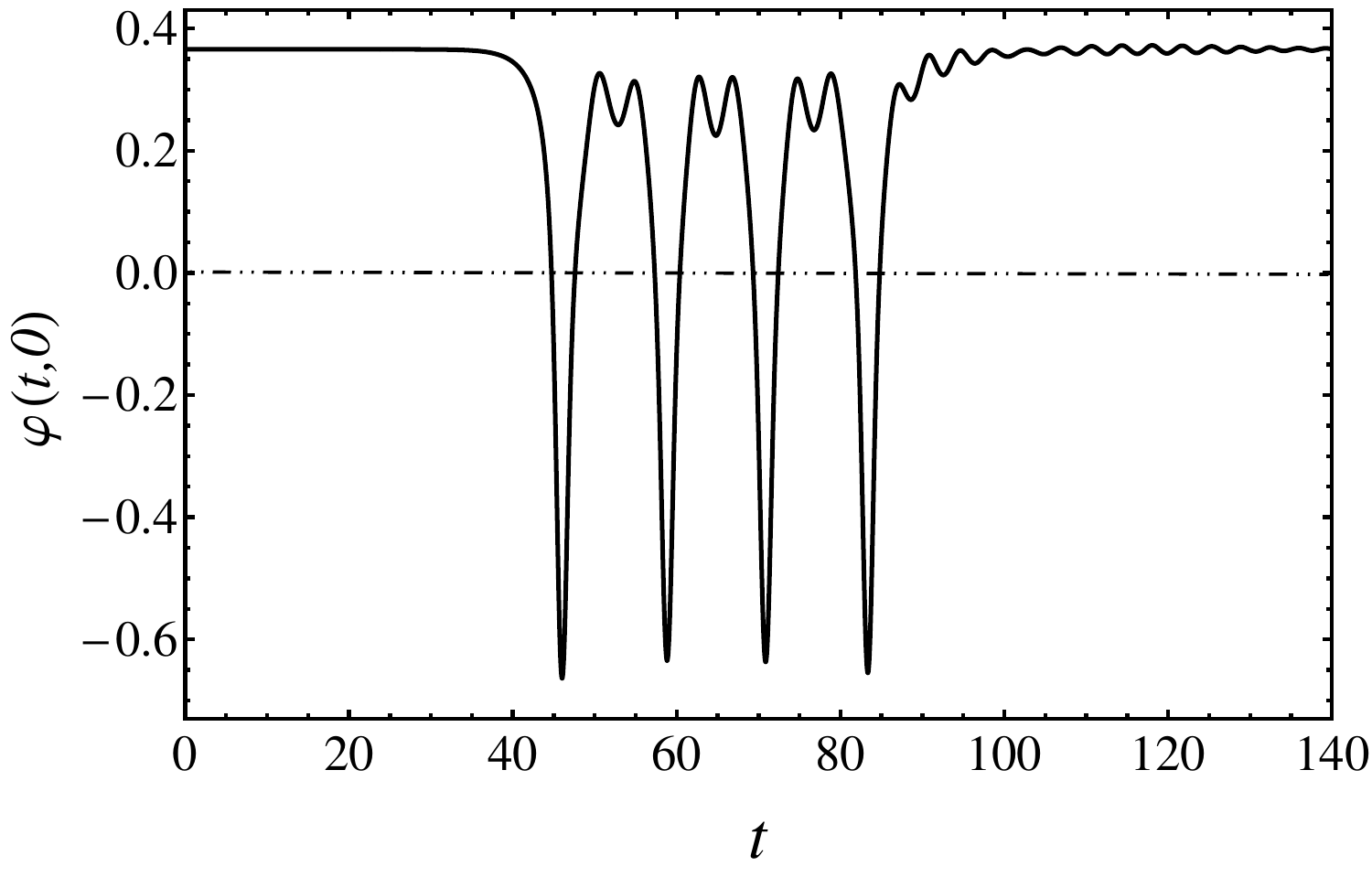}}
\end{minipage}
\caption{Four-bounce at $v_{\scriptsize \mbox{in}}= 0.2568$. {\bf Left panel} --- space-time evolution. {\bf Right panel} --- the profile of $\varphi(t,0)$.}
\label{fig:window4_2568}
\end{figure}

As mentioned above, in our calculations we also observed three-bounces and four-bounces
--- in these cases the kinetic energy is (partly) restored only after three and four collisions,
respectively. Typical space-time pictures of these processes are shown in
Figs.~\ref{fig:window3_3012} and \ref{fig:window4_2568}.
Another resonance feature, observed previously in kink-kink collisions
in the double sine-Gordon~\cite{GaKuPRE,campbell} as well as in the modified sine-Gordon model~\cite{peyrard}, is quasi-resonances.
They occur when the energy that is pumped back into the translational mode
during the second collision is not enough for the two kinks to escape.
The two kinks merely come some large but finite distance apart and then collide again, eventually merging into the bion. This situation shows
as a peak on the plot of $T_{23}$, the time between the second and
the third collisions, as a function of the initial velocity
$v_\mathrm{in}$.
In some models, quasi-resonances have been shown to replace some of the escape windows \cite{GaKuPRE},
which is not unexpected as both these phenomena are due to the resonant energy exchange.
Our calculation has not shown any clear signs of quasi-resonances occurring in the
considered variant of the $\varphi^8$ model.

\section{\label{sec:level5} Summary and discussion}

We studied excitation spectra of the $\varphi^8$ field model, choosing
three different sets of the model's self-interaction potential.
We found that some of the kinks that arise in those variants of
the $\varphi^8$ model have vibrational excitation modes, which
points at the possibility for resonance phenomena to occur in
low-energy kink-kink scattering. To illustrate that relation
between the excitation of a solitary kink and the kink-kink scattering,
we performed numerical modelling of the latter process, using one
of the considered kinks that has a vibrational excitation, and the corresponding antikink.
We showed that there are two collision regimes, depending on the
initial velocity of the two colliding kinks. Namely,
the kinks bounce off each other inelastically at
$v_{\scriptsize \mbox{in}}\ge v_{\scriptsize \mbox{cr}}$, whereas
$v_{\scriptsize \mbox{in}}<v_{\scriptsize \mbox{cr}}$
results in the formation of a bound state of two kinks, the bion.
Furthermore, we demonstrated the existence of escape windows
--- intervals of initial velocities in the domain $v_{\scriptsize \mbox{in}}<v_{\scriptsize \mbox{cr}}$ where, even though these velocities
are below the critical speed, the kinks escape to infinities
after two, three, four etc.\ collisions. Our analysis of two-bounce
escape windows shows that their positions and structure are well described by the phenomenological
resonance condition, with the corresponding resonance frequency being close to the frequency
of the kink's vibrational mode. This confirms the resonant energy exchange between the kink's
translational mode and its vibrational mode (or, more precisely, a localised excitation of
the kink-kink system that is close to the vibrational mode of a solitary kink) as the
driving mechanism that leads to the occurrence of escape windows in kink-kink collisions
in that sector of the $\varphi^8$ model.

Note that the value of the critical speed $v_{\scriptsize \mbox{cr}}\simeq 0.3160$
that results from our analysis of collisions between the kinks $(-a,a)$ and
$(a,-a)$ corresponds to the particular choice of the model parameters
$a$ and $b$. One can expect that the critical speed will be a function of these parameters; this distinguishes the $\varphi^8$ model from such models
as $\varphi^4$ and $\varphi^6$, where there are no free parameters in the model
(or, more strictly, they can be eliminated by appropriate rescaling of the field
and the space-time coordinates). The critical speeds in these latter models are
therefore fixed, namely, $v_{\scriptsize \mbox{cr}}\simeq 0.2598$ in the $\varphi^4$
model \cite{aek01}, whereas the $\varphi^6$ model \cite{GaKuLi} has two different critical speeds
(there are two different topological configurations of colliding kinks
where the formation of a bion is possible and hence the critical speed can be
defined), $v_{\scriptsize \mbox{cr}}\simeq 0.289$ for the colliding kinks $(0,-1)$
and $(-1,0)$ and $v_{\scriptsize \mbox{cr}}\simeq 0.045$ for the kinks $(-1,0)$
and $(0,-1)$. In this connection, a more illustrative example is that of
the modified sine-Gordon model \cite{peyrard},
where the model potential has a single parameter $r$ such that at $r\to 0$
one recovers the regular sine-Gordon model (which is fully integrable and
formally has $v_{\scriptsize \mbox{cr}}= 0$). The modified sine-Gordon model
yields the following critical speeds: $v_{\scriptsize \mbox{cr}}\simeq 0.112$ at
$r=0.05$, $v_{\scriptsize \mbox{cr}}\simeq 0.234$ at $r=0.1$,
$v_{\scriptsize \mbox{cr}}\simeq 0.337$ at $r=-0.5$. One can clearly see that
the critical speed depends on the model parameter and that its value increases
as the model moves away from the integrable limit. As far as the $\varphi^8$
model is concerned, it is probably safe to state that the critical velocity
in the collisions that we studied can be made as low as that of the $\varphi^4$ model.
Indeed, one could consider a situation when $b$ is chosen much larger than $a$.
The relevant model potential, Eq.~(\ref{eq:potphi8}),
would in this case be very close to that of the $\varphi^4$ model
as long as $\varphi\sim a$,
leading to the dynamics of collisions between the kinks $(-a,a)$ and $(a,-a)$
being close to that of kink-antikink collisions in the $\varphi^4$ model.

Finally, we would like to reflect on several issues that have been
left out of the scope of this work but represent interest
for future research.

\begin{enumerate}
\item Some of the kinks of the $\varphi^8$ model, corresponding to
particular choices of the model parameters, can have power-law,
rather than exponential, asymptotic at the spatial infinity~\cite{khare}.
Studying these kinks along the lines of our work would be a natural
extension of this work.

\item Resonant energy exchange in kink-kink collisions can
also occur when the corresponding solitary kinks do not have
vibrational excitations. For instance, two colliding kinks
can have a collective vibrational mode, even though each of the
two solitary waves does not. This has been shown to happen
in the $\varphi^6$ and the double-sine-Gordon models~\cite{dorey01,GaKuPRE}.
This situation, apparently,
can take place in some variants of the $\varphi^8$ model.

\item We think (in agreement with the authors of ref.~\cite{khare}) that the collective coordinate
method could
be productively applied to studying kink-kink interaction
in this model. It is worth noting that kinks with power-law
asymptotic can apparently result in power-law asymptotic of
the collective potential, which is a new feature that 
does not occur in other field models, such as, e.g.,
$\varphi^4$ and $\varphi^6$.
\end{enumerate}

\section*{Acknowledgements}
The authors are very grateful to Prof.~A.~E.~Kudryavtsev for his interest to their work and
for enlightening discussions. This work was supported in part by the Russian Federation Government under grant No.~NSh-3830.2014.2. V.~A.~Gani acknowledges support of the Ministry of Education and Science of the Russian Federation, Project No.~3.472.2014/K. M.~A.~Lizunova and V.~Lensky thank the ITEP support grant for junior researchers. M.~A.~Lizunova also gratefully acknowledges financial support from the Dynasty Foundation.

\end{document}